\documentclass[preprint,12pt]{elsarticle}

\usepackage{amssymb}
\usepackage{amsmath}
\usepackage{float}
\usepackage{multirow}
\usepackage{float} %
\usepackage{booktabs} %

\begin{document}

\begin{frontmatter}

\title{\textbf{MedVQA-TREE: A Multimodal Reasoning and Retrieval Framework for Sarcopenia Prediction}}

\author[inst1,inst2]{Pardis Moradbeiki}
\ead{pamo@mmmi.sdu.dk}

\author[inst1]{Nasser Ghadiri}
\ead{nghadiri@iut.ac.ir}

\author[inst1]{Sayed Jalal Zahabi\corref{cor1}}
\ead{zahabi@iut.ac.ir}

\author[inst2]{Uffe~Kock~Wiil}
\ead{ukwiil@mmmi.sdu.dk}

\author[inst3]{Kristoffer Kittelmann Brockhattingen}
\ead{Kristoffer.K.Brockhattingen@rsyd.dk}

\author[inst2]{Ali Ebrahimi}
\ead{aleb@mmmi.sdu.dk}

\affiliation[inst1]{
    organization={Department of Electrical and Computer Engineering, Isfahan University of Technology},
    city={Isfahan},
    country={Iran}
}
\affiliation[inst2]{
    organization={SDU Health Informatics and Technology, The Maersk Mc-Kinney Moller Institute, University of Southern Denmark},
    city={Odense},
    country={Denmark}
}

\affiliation[inst3]{
    organization={Geriatric Research Unit, Department of Clinical Research, University of Southern Denmark},
    city={Odense},
    country={Denmark}
}

\begin{abstract}
Accurate sarcopenia diagnosis via ultrasound remains challenging due to subtle imaging cues, limited labeled data, and the absence of clinical context in most models. We propose MedVQA-TREE, a multimodal framework that integrates a hierarchical image interpretation module, a gated feature-level fusion mechanism, and a novel multi-hop, multi-query retrieval strategy. The vision module includes anatomical classification, region segmentation, and graph-based spatial reasoning to capture coarse, mid-level, and fine-grained structures. A gated fusion mechanism selectively integrates visual features with textual queries, while clinical knowledge is retrieved through a UMLS-guided pipeline accessing PubMed and a sarcopenia-specific external knowledge base.  MedVQA-TREE was trained and evaluated on two public MedVQA datasets (VQA-RAD and PathVQA) and a custom sarcopenia ultrasound dataset. The model achieved up to 99\% diagnostic accuracy and outperformed previous state-of-the-art methods by over 10\%. These results underscore the benefit of combining structured visual understanding with guided knowledge retrieval for effective AI-assisted diagnosis in sarcopenia.
\end{abstract}


\end{frontmatter}
\begin{abstract}
Accurate sarcopenia diagnosis via ultrasound remains challenging due to subtle imaging cues, limited labeled data, and the absence of clinical context in most models. We propose MedVQA-TREE, a multimodal framework that integrates a hierarchical image interpretation module, a gated feature-level fusion mechanism, and a novel multi-hop, multi-query retrieval strategy. The vision module includes anatomical classification, region segmentation, and graph-based spatial reasoning to capture coarse, mid-level, and fine-grained structures. A gated fusion mechanism selectively integrates visual features with textual queries, while clinical knowledge is retrieved through a UMLS-guided pipeline accessing PubMed and a sarcopenia-specific external knowledge base. MedVQA-TREE was trained and evaluated on two public MedVQA datasets (VQA-RAD and PathVQA) and a custom sarcopenia ultrasound dataset. The model achieved up to 99\% diagnostic accuracy and outperformed previous state-of-the-art methods by over 10\%. These results underscore the benefit of combining structured visual understanding with guided knowledge retrieval for effective AI-assisted diagnosis in sarcopenia.
\end{abstract}





\section{\textbf{Introduction} }
\label{sec1}
Sarcopenia, a major giant of geriatric medicine, is characterized by the progressive and functional loss of skeletal muscle, induced by a loss of mass, strength and endurance \cite{dupont2018revised} The condition is commonly associated with aging, and factors such as malnutrition, physiological “wear and tear” due to the burden of chronic disease (multimorbidity), and reduced physical activity can exacerbate the condition or in combination with one and another induce it. The heterogeneous patients that suffer from the condition challenges its clinical presentation which poses great obstacles for a patient-centered diagnosis and personalized treatment \cite{cho2022review, soria2024ai, yu2021relationship, zuo2023sarcopenia}.

Imaging techniques like dual-energy X-ray absorptiometry (DXA) and bioelectrical impedance analysis (BIA) provide estimates of muscle mass, while Computed Tomography (CT) and Magnetic resonance imaging (MRI) offer insights into muscle composition and fat infiltration. However, their high cost, limited accessibility, and radiation concerns limit widespread use \cite{kim2025comparison, wu2024correlation}. Radiomics-based analysis enables extraction of high-dimensional features from medical images but still faces issues like lack of external validation and limited functional integration \cite{tomaszewski2021biological, kocak2020radiomics}.

To address these limitations, AI-driven multimodal methods are gaining traction, combining imaging with biomechanical data from wearable devices (e.g., Inertial Measurement Units (IMUs), smart insoles, pose estimation) to enable continuous, real-world monitoring of sarcopenia \cite{kim2023assessing, kim2023sarcopenia}. Yet, challenges remain in longitudinal tracking, standardization, and heterogeneous data integration \cite{chen2021towards, islam2022fully, sakai2021machine, caputo2025towards, aznar2024gait}. Among imaging modalities, ultrasound has gained significant attention due to its non-invasive, portable, affordable, and radiation-free nature, enabling real-time assessment of both muscle quantity (e.g., thickness, Cross-Sectional Area (CSA)) and quality (e.g., echogenicity, pennation angle, fascicle length) \cite{kim2023assessing, kim2023sarcopenia, smorchkova2022sarcopenia}. Its suitability for bedside use and community-level screening makes it particularly valuable for diagnosing and monitoring sarcopenia in older adults. However, operator dependency and limited sensitivity to subtle muscle changes necessitate advanced, multi-level image analysis to reduce observer bias and improve diagnostic reliability. Moreover, integrating sonographic features with clinical data such as comorbidities, tests of physical functioning, and laboratory results can provide a more holistic and accurate diagnosis. To support the clinical adoption of ultrasound in muscle assessment, collaborative efforts such as SARCUS (SARCopenia through UltraSound), an European Geriatric Medicine Society initiative, are working to standardize acquisition protocols, determine key diagnostic parameters, and collect population-based reference data \cite{perkisas2019sarcus}. This initiative aims to close critical knowledge gaps by identifying the most clinically relevant ultrasound biomarkers and integrating them into comprehensive geriatric assessments. Such international collaborations are essential to establish consensus protocols and promote the use of muscle ultrasound at the bedside, where timely and accessible diagnosis of sarcopenia is urgently needed.

Recent advances in AI, especially deep learning and multimodal data fusion, offer promising tools to enhance image interpretation, integrate clinical knowledge graphs, and support decision-making in complex conditions \cite{smorchkova2022sarcopenia, kim2023sarcopenia, behboodi2024deepsarc}. Despite success in fields like oncology and neurology, AI applications in sarcopenia are still limited due to small, heterogeneous datasets, lack of domain-specific pipelines, inconsistent imaging protocols, and limited interpretability \cite{caputo2025towards, ye2023development}. To address these gaps, studies have explored predictive models combining clinical indicators, image-derived features, and multi-omics data \cite{lu2022multi}. A novel direction involves multimodal large language models (MLLMs), which jointly analyze medical images and clinical texts to answer diagnostic questions. While powerful in general domains, their clinical deployment in areas like sarcopenia remains challenged by data scarcity and domain adaptation needs.

Most current systems use either general-purpose (e.g., VGG, ResNet, BERT) \cite{lee2021clinical} or medical-domain (e.g., MedSLIP \cite{fan2024medslip}, FetalCLIP \cite{maani2025fetalclip}) models, but often fail to capture fine-grained anatomical details or interpret nuanced clinical language, especially in ultrasound reports with specialized or multilingual terminology \cite{ge2025ultrasound}. While some architectures like MeDSLIP improve anatomical awareness and others incorporate knowledge bases to enhance text understanding, many still lack task-specific optimization and robust inference capabilities \cite{yi2024clinical}.

A key limitation in MLLMs is the weak alignment between image and text modalities, often leading to hallucinations or clinically irrelevant outputs. For instance, many models over-rely on textual priors and may generate outputs inconsistent with visual evidence due to autoregressive next-token prediction mechanisms. Approaches like LoGra-Med improve semantic alignment through triplet supervision (image caption dialogue), and retrieval-augmented generation (RAG) with visual grounding can reduce hallucinations by anchoring outputs in real content. However, integration of retrieved knowledge with visual data and user queries remains suboptimal in structured clinical workflows \cite{chu2025reducing, nguyenenriched, toro2024dynamic, nguyenlogra}.

Integrating external knowledge via RAG offers promise, as seen in frameworks like Bailicai, which leverage medical ontologies to enhance factuality. Yet, these methods often depend on hand-crafted queries, suffer from noisy document retrieval, and lack effective fusion mechanisms. Addressing these issues requires improved strategies for retrieval, alignment, and grounding, specifically tailored to medical imaging and clinician interaction.

Therefore, the development of clinically reliable MLLMs demands novel strategies for semantic alignment, domain adaptation, and knowledge integration, particularly when dealing with unstructured inputs such as ultrasound images and narrative clinical text in underrepresented conditions like sarcopenia. To address the aforementioned challenges, we hypothesize that a lightweight multimodal deep learning model, which integrates high-resolution transverse and longitudinal ultrasound scans with structured clinical data (including age, body mass index (BMI), and Short Physical Performance Battery (SPPB) scores), can effectively improve the diagnostic performance for sarcopenia. To improve interpretability and generalization in a low-data setting, we employ compact transformer-based modules for multi-scale feature encoding. The key contributions of this study are as follows:

\begin{enumerate}
    \item A hierarchical tree-based framework tailored for sarcopenia-aware visual reasoning: We introduce Medical Visual Question Answering using Tree-based Reasoning (MedVQA-TREE), a multi-level framework specifically designed for sarcopenia diagnosis from ultrasound images. Unlike existing models that treat visual features as flat embeddings, our approach mimics clinical reasoning by performing coarse-to-fine visual reasoning. high-level anatomical classification, region-specific segmentation (e.g., muscle borders), and graph-based spatial modeling to capture relative positions and interaction of anatomical structures. This hierarchical pipeline reflects clinical practice and addresses the challenge of limited visual saliency in muscle ultrasound.

    \item Integration of RAG with structured medical knowledge: Our framework incorporates a UMLS-guided RAG module that dynamically accesses evidence from PubMed and a curated sarcopenia-specific knowledge base. Unlike standard RAG systems that typically rely on single-turn queries, our method leverages multi-query and multi-hop retrieval strategies to enhance conceptual coverage and reasoning depth. By generating multiple semantically distinct queries based on structured patient features (including age, BMI, muscle strength, comorbidities), the system retrieves evidence across diverse clinical dimensions. Furthermore, the multi-hop mechanism enables the model to identify indirect medical connections such as the link between obesity, inflammation, and muscle degradation which are especially relevant in complex syndromes like sarcopenia. This retrieval process is grounded in standardized medical ontologies, improving factual accuracy, minimizing hallucination, and ensuring clinically reliable response generation.

    \item A relation-aware gated fusion mechanism: We propose a novel mechanism that selects the most relevant level of visual abstraction (coarse, mid, or fine) based on the input question using a question-guided gate selector. This selected representation is then passed through a Low-Rank Adaptation (LoRA) module for cross-modal fusion with the clinical knowledge. This design improves semantic alignment, reduces unnecessary noise from unrelated visual levels, and enhances interpretability in medical VQA tasks.

    \item Comprehensive evaluation across public and domain-specific datasets: We validate our model on two widely used medical VQA datasets (VQA-RAD \cite{lau2018dataset}, PathVQA \cite{he2020pathvqa}) as well as a newly collected ultrasound-based sarcopenia dataset.

\end{enumerate}
\section{\textbf{Materials and Methods} }
\label{sec2}

In this section, a multi-modal framework is introduced to address the challenges associated with sarcopenia diagnosis and analysis. The following subsections describe the procedures for data collection, preprocessing, feature extraction, model development, and evaluation. The proposed framework comprises environment three main modules, each targeting a distinct aspect of the problem: 

\textit{\textbf{Hierarchical Ultrasound Image Interpretation:}} This module employs a three-level visual encoder to capture coarse, mid-level, and fine-grained features from ultrasound images, focusing on muscle structure and tissue characteristics critical for sarcopenia diagnosis. Image preprocessing techniques are tailored to each level of the hierarchical tree structure to ensure consistent input quality, as described in Section 2.2. Section 2.3 then outlines the feature extraction process, and in Section 2.3.1, we detail the image feature extraction at three levels: Level 1 – High-Level Global Features (Coarse Grain), Level 2 – Regional Feature Extraction via Segmentation (Mid Grain), and Level 3 – Fine-Grained Spatial Graph Features (Fine Grain). Different architectures are applied at each granularity level to generate context-aware representations, enhancing the model’s ability to interpret clinically relevant visual patterns.

\textit{\textbf{UMLS-Guided Biomedical Knowledge Retrieval:}} This module implements a dynamic retrieval strategy to construct a sarcopenia-specific external knowledge base using multi-query and multi-hop retrieval over PubMed and UMLS clinical ontologies. This process includes two main steps: Pre-processing and Feature extraction. In the preprocessing phase, structured patient variables (e.g., age, BMI, SPPB) are transformed into UMLS-enriched clinical concepts and expanded into multiple semantically diverse queries, as described in Section 2.2. The knowledge base construction process is then detailed, including the integration of UMLS, the generation of multi-hop queries, and the retrieval of relevant documents from PubMed. Retrieved content is segmented and indexed using a sarcopenia-specific Latent Dirichlet Allocation (LDA) model to ensure contextual alignment. Section 2.3 further elaborated on feature extraction. Section 2.3.2: Numeric Feature Extraction explains how structured patient data is used to guide query generation. Section 2.3.3: Textual Feature Extraction describes the Retrieval-Augmented Generation (RAG) mechanism, highlighting how top-ranked sentences are retrieved and encoded to provide clinically relevant textual features during inference.

\textit{\textbf{Question-Guided Gated Fusion:}} This module introduces a question-guided gated fusion mechanism that selectively integrates multi-modal features both visual (from different abstraction levels) and textual (from retrieved clinical knowledge) to enhance diagnostic accuracy and semantic alignment in the final prediction. The fusion process is dynamically modulated by the clinical question, allowing the model to attend to the most relevant features of any level based on contextual cues: Model Development, we mathematically formalize this fusion architecture, detailing how the gating mechanism is applied across feature levels and how the final joint representation is computed for classification, as described in Section 2.4. This module plays a critical role in aligning the retrieved evidence and visual cues with the intent of the question. 

Figure 1 provided a comprehensive overview of the study’s methodology, encompassing all stages of the framework. Each subfigure corresponds to specific model components, described in detail across Sections 2.2–2.4: (a) Ultrasound images are first categorized into transverse or longitudinal views using a lightweight orientation-specific classifier. Subsequently, 400 domain-optimized global features are extracted to capture coarse-grain anatomical patterns. (b) Region-based features are extracted via a SAM-based segmentation model, which identifies and isolates anatomically relevant muscle areas. From the top-S most informative regions, 400-dimensional vectors are computed to facilitate mid-grain analysis of localized muscle degradation. (c) Fine-grained spatial relationships between anatomical structures are modeled using a superpixel-based graph. Here, node embeddings encode localized structural dependencies, enabling the detection of subtle patterns associated with early-stage muscle degradation. → These processes ( a, b and c) are detailed in Section 2.2.1 (Image Preprocessing) and Section 2.3.1 (Visual Feature Extraction: Hierarchical – Coarse, Mid, Fine Grain). (d) Structured clinical variables such as age, sex, BMI, and SPPB are encoded via neural projection layers into 400-dimensional latent representations that capture both statistical distributions and biomedical semantics. → Described in Section 2.2.2 (Numeric Data Preprocessing) and Section 2.3.2 (Numeric Feature Extraction). (e) An adaptive gated fusion mechanism dynamically selects the most informative modality-specific representation, enabling effective integration of visual and clinical information through a LoRA-enhanced network for downstream classification. → Detailed in Section 2.4 (Model Development). (f) A semantic retrieval module converts patient-specific clinical features into UMLS-guided multi-query PubMed searches. Retrieved evidence is filtered and refined through multi-hop retrieval, supporting knowledge enrichment tailored to sarcopenia diagnosis. → Further discussed in Section 2.2.3 (Textual Query Processing) and Section 2.3.3 (Textual Feature Extraction). (g) Retrieved sentences are topic-filtered using LDA and embedded via biomedical language models such as BioBERT and PubMedBERT. The top-C contextually aligned sentences are integrated with imaging and clinical features through a RAG-based architecture, enabling personalized, knowledge-guided representations that enhance both interpretability and diagnostic accuracy. → Fully described in Section 2.3.3 (Textual Feature Extraction) and Section 2.4 (Model Development).
\begin{figure}
    \centering
    \scriptsize
    \includegraphics[width=1.0\linewidth]{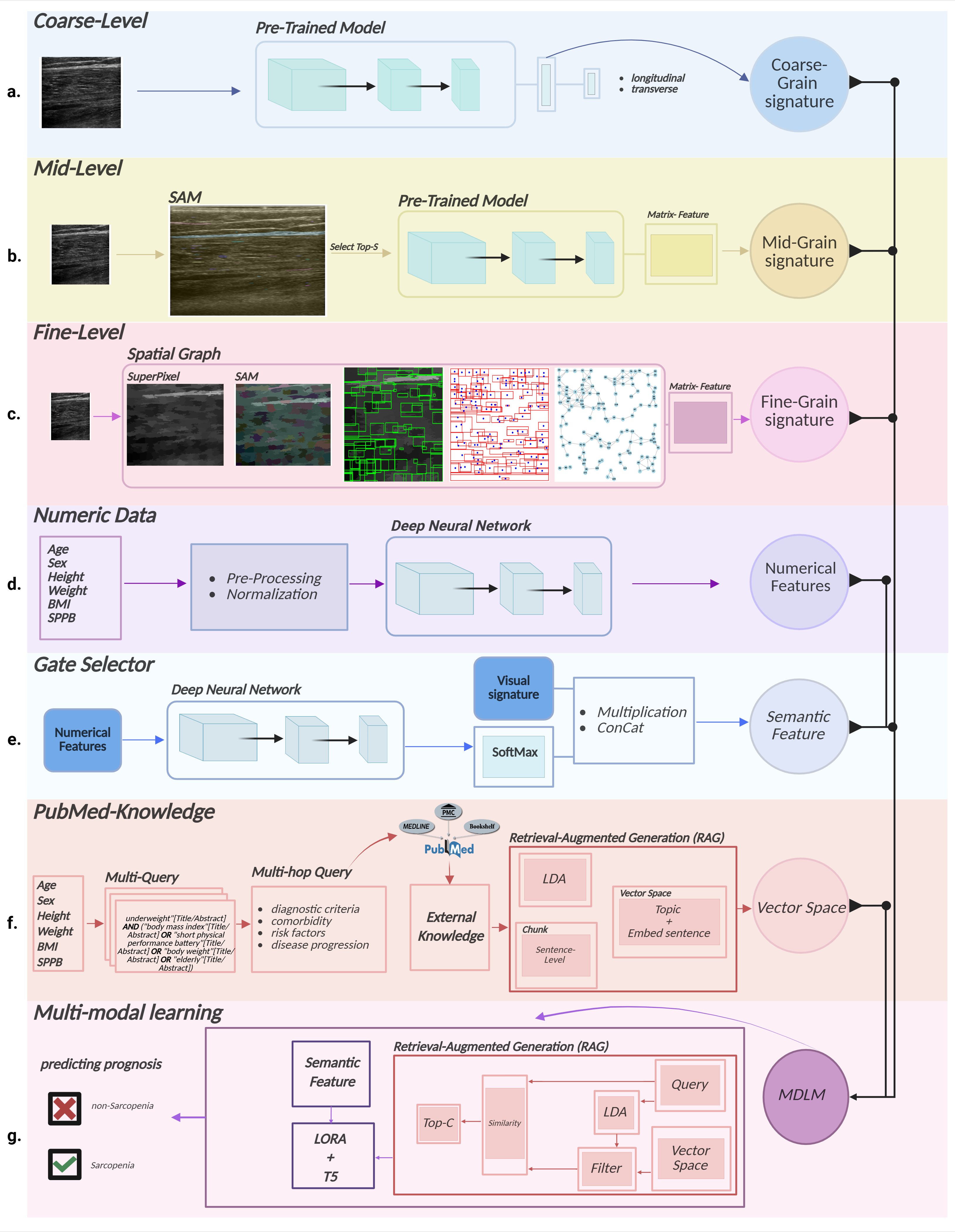}
    \caption{Overview of the proposed multi-modal framework for sarcopenia.}
    \label{fig:enter-label}
\end{figure}
\subsection{\textbf{Data collection} }
\label{sec2}
To evaluate the disease-specific prediction capability of our proposed model, MedVQA-TREE was first assessed on a custom ultrasound dataset focusing on sarcopenia. We then applied our method to two publicly available MedVQA datasets as downstream tasks, namely VQA-RAD and PathVQA, to assess the generalizability and domain-specific effectiveness of the model. A description of each dataset follows.

\textit{\textbf{Sarcopenia Ultrasound Dataset (Proprietary):}} 
The dataset comprises multimodal patient data, including both structured clinical variables and unstructured ultrasound imaging, aimed at supporting sarcopenia classification. A total of 24 patients were enrolled and categorized as either sarcopenic or non-sarcopenic (control), classified according to the European Working Group on Sarcopenia in older people 2 evidence-based criteria \cite{dupont2018revised}. The structured data includes demographic and physical attributes such as age, sex, height, weight, BMI, the number of self-reported falls, and the Short Physical Performance Battery (SPPB) score, a validated tool for assessing lower limb strength and fall risk in older adults. The cohort consisted of 73\% female and 27\% male participants, consistent with the higher prevalence of sarcopenia among women. The imaging component contains 3,474 grayscale ultrasound images captured in both transverse (1,350 images: 458 sarcopenic, 892 control) and longitudinal (962 images: 318 sarcopenic, 644 control) planes, focusing on the Rectus Femoris (RF) muscle at the anterior thigh. Images were acquired at two high resolutions 1552×970 and 1164×873 pixels allowing detailed assessment of muscle structure and texture relevant to sarcopenia detection. Overall, 30.53\% of the images were labeled as sarcopenic and 69.47\% as non-sarcopenic. Stratified by gender, male patients contributed 763 sarcopenic and 1,240 non-sarcopenic samples, while female patients contributed 297 sarcopenic and 1,172 non-sarcopenic samples. This curated dataset supports the development and evaluation of multimodal AI models for early sarcopenia diagnosis.

\textit{\textbf{VQA-RAD Dataset:}}
VQA-RAD is a radiology-focused VQA dataset comprising 315 manually curated medical questions paired with 515 radiology images (X-ray, CT, MRI). Each question falls into one of the predefined types (Yes/No, Open-ended, Counting) and is linked to relevant anatomical or pathological contexts. Despite its small size, it is widely used for evaluating fine-grained medical reasoning in low-data regimes.

\textit{\textbf{PathVQA Dataset:}}
PathVQA includes 4,998 image question answer triplets from pathology slides. It covers detailed diagnostic scenarios across multiple organs, emphasizing both visual detail and factual grounding. It serves as a high-fidelity benchmark for knowledge-augmented visual diagnosis.
\subsection{\textbf{Pre-processing} }
\label{sec2}
The custom sarcopenia dataset was imbalanced, with 69.47\% of the samples labeled as non-sarcopenic and 30.53\% as sarcopenic. This class imbalance was also reflected across sex subgroups. Among male patients, 1,240 samples were non-sarcopenic and 763 were sarcopenic, while among female patients, 1,172 samples were non-sarcopenic and 297 were sarcopenic.
All ultrasound images and associated clinical variables were collected as part of a custom sarcopenia dataset designed to reflect diverse anatomical views and varying stages of muscle degradation. For each patient, multiple longitudinal and transverse ultrasound images were included, capturing distinct cross-sectional muscle features. No filtering was applied to reduce the number of images per patient; instead, all available images were retained to enrich the visual context and allow the model to leverage complementary spatial perspectives.
\subsubsection{\textbf{Image Preprocessing} }
\label{sec2}
All ultrasound images were normalized and resized to a fixed resolution to ensure consistency across the dataset. Several preprocessing techniques were investigated, and their comparative performance is presented in Section 3 (Results).
\subsubsection{\textbf{Numeric Data Preprocessing} }
\label{sec2}
Structured clinical variables including age, sex, height, weight, BMI, and SPPB score were preprocessed via two complementary pathways to support both multimodal fusion and external knowledge retrieval:
\begin{enumerate}
\item Embedding Path (Feature Integration)
To integrate numerical data with image-derived features, clinical variables were encoded into vector representations. Specifically, categorical variables, such as sex, were encoded using one-hot vectors or trainable embedding layers. Continuous variables, including BMI, weight, height, and SPPB score, were normalized using min-max scaling to ensure consistency and avoid feature dominance during fusion.
Several preprocessing strategies were explored, and the optimal configurations were selected based on empirical evaluations, as discussed in the Results section.
\item Knowledge Retrieval Path (Semantic Enrichment)
To supplement the model with domain-specific evidence, we developed a structured pipeline that translates patient-level features into semantically enriched PubMed queries. This process leverages the UMLS API for synonym expansion and concept normalization.
For each patient, key numerical or categorical features such as BMI, SPPB, age, and gender were converted into standardized clinical terms. These terms were then mapped to UMLS Concept Unique Identifiers (CUIs) and expanded into semantically related concepts (e.g., "muscle weakness" → "sarcopenia", "frailty", "muscle atrophy").  Each patient yielded 10–20 such queries. Using the Entrez API \cite{NCBI2009}, we retrieved the top 10 abstracts per query. These abstracts were further filtered using UMLS-based relevance scoring and LDA topic modeling, ensuring that the most clinically relevant and semantically aligned information was retained.
Unlike conventional LLMs or standard RAG systems that rely on pre-indexed corpora or static knowledge bases, our approach introduces a domain-specific, on-demand biomedical retrieval framework. Rather than depending on potentially outdated or hallucinated content, the model dynamically retrieves up-to-date clinical evidence from trusted sources such as PubMed.
While several recent RAG variants such as DRAGON-AI for ontology-guided retrieval \cite{toro2024dynamic}, hybrid three-stage medical RAG architectures \cite{yang2025dual}, and Dynamic Retrieval-Augmented Generation (DRAG) \cite{xiong2024improving, amugongo2025retrieval} support live or hybrid retrieval workflows, they have not been specifically designed or validated for ultrasound-based sarcopenia diagnosis. In contrast, our method is purpose-built for this domain, introducing a tailored on-demand retrieval mechanism aligned with clinical diagnostic requirements and grounded in high-quality biomedical knowledge.
To implement this, we generate clinically enriched queries using structured patient-level variables, which are semantically expanded into medical concepts via UMLS. These queries span multiple clinical dimensions including diagnostic criteria, comorbidities, risk factors, and disease progression. Retrieved PubMed documents are segmented into sentences and filtered through topic modeling using a sarcopenia-specific LDA model to ensure contextual relevance. The resulting content is stored in an external knowledge base dedicated to sarcopenia and used during inference to enhance model reasoning. Technical details of the pipeline are provided in the following sections.
\end{enumerate}
\subsubsection{\textbf{Textual Query Processing} }
\label{sec2}
All queries generated from structured clinical data used for retrieving text from biomedical knowledge sources underwent standardized preprocessing prior to feature extraction. These preprocessing steps were designed to enhance the relevance and consistency of retrieved documents and ensure compatibility with downstream text encoders. Specifically:
\begin{itemize}
    \item All text was converted to lowercase to ensure case-insensitive matching.
    \item Stop words and punctuation marks were removed to eliminate non-informative tokens.
    \item Queries were then tokenized using a biomedical-aware tokenizer, optimized for domain-specific terminology and abbreviations commonly found in clinical and research texts.
\end{itemize}
\subsection{\textbf{Feature extraction } }
\label{sec2}
To enable accurate visual queries answering for sarcopenia diagnosis from multimodal clinical inputs, it is essential to extract rich, semantically meaningful features from images, structured clinical variables, and external knowledge sources. Our model, MedVQA-TREE, incorporates a multi-path feature extraction strategy tailored to the distinct characteristics of each modality: ultrasound images, numerical data, and medical text.
\subsubsection{\textbf{Visual Feature Extraction (Hierarchical: Coarse, Mid, and Fine Grain) } }
\label{sec2}
To address the challenges of analyzing medical images such as sarcopenia, we developed a three-level hierarchical feature extraction module tailored for visual data. This architecture progressively extracts features from global to fine-grained anatomical detail, enabling adaptive processing based on the complementary value of clinical data.

\textbf{Level 1 – High-Level Global Features (Coarse Grain):}
At the first level, each ultrasound image is classified into transverse or longitudinal views using a lightweight classifier (1 dense layer, 10 neurons), trained specifically on image orientation. Subsequently, 400 abstract global features are extracted from each image. This level aligns with the common approach in previous studies like \cite{zhang2023biomedclip} that extract global visual features using pretrained networks, but it is domain-optimized and lighter to reduce computation and increase efficiency for ultrasound data. 

\textbf{Level 2 – Regional Feature Extraction via Segmentation (Mid Grain): }
The second level utilizes Segmentation Anything Model (SAM) \cite{kirillov2023segment}, a pre-trained medical segmentation model, to identify relevant anatomical muscle and tissue regions. From the segmented mask, the top-S most informative regions are selected (details of n and selection criteria in Section 3.2.1. ). For each selected region, 400 feature vectors are extracted, forming an S × 400 matrix. This step refines the feature space by focusing on contextual muscle structure and local variations. As shown in Figure 2, the SAM-based model effectively segments key regions and extracts mid-level features from the most informative areas (image selected of VQA-RAD dataset). 
\begin{figure}
    \centering
    \scriptsize
    \includegraphics[width=1\linewidth]{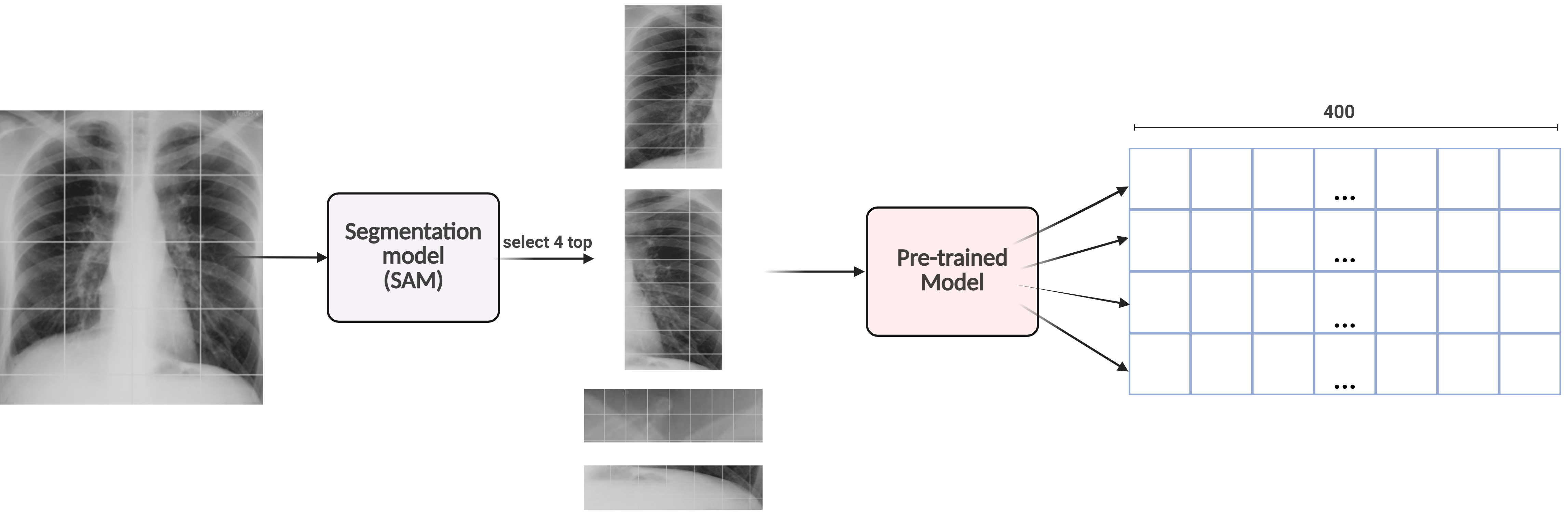}
    \caption{Region-Based Feature Extraction via SAM Segmentation (image selected of VQA-RAD dataset)}
    \label{fig:enter-label}
\end{figure}
\textbf{Level 3 – Fine-Grained Spatial Graph Features (Fine Grain): }
At the third level, we construct a graph-based representation to capture fine-grained anatomical and spatial relationships among localized muscle regions. The process begins by oversegmenting the ultrasound image into superpixels using a segmentation algorithm (create-superpixel-graph), generating visually coherent regions that serve as initial building blocks for the graph. Each superpixel region is then passed through SAM, a pretrained segmentation model, to extract the top-S most informative anatomical areas based on clinical saliency and structural clarity. For each selected region, centroid coordinates are computed and used to define nodes in the graph. These nodes are connected by edges based on spatial proximity, geometric continuity, and anatomical adjacency, forming a "graph pixel map" where each node corresponds to a clinically meaningful muscle segment. Each node is enriched with metadata such as its bounding box and position, enabling spatially grounded modeling of muscle structure. The resulting graph is processed using a lightweight graph embedding module (generate-node-embeddings), which produces dense vector representations for each node based on its topological role and spatial location. These embeddings are aggregated into an S × 400 matrix, and a PCA-inspired fixed-size transformation (to-fixed-size-vector) is applied to form a unified 400-dimensional feature vector. This encoding preserves localized anatomical cues while ensuring compatibility with downstream reasoning modules. As visualized in Figure 3, this step allows the model to reason over the spatial configuration of muscle regions, effectively capturing subtle visual patterns associated with sarcopenia, such as texture degradation, regional asymmetry, and localized atrophy.
\begin{figure}
    \centering
    \scriptsize
    \includegraphics[width=1.0\linewidth]{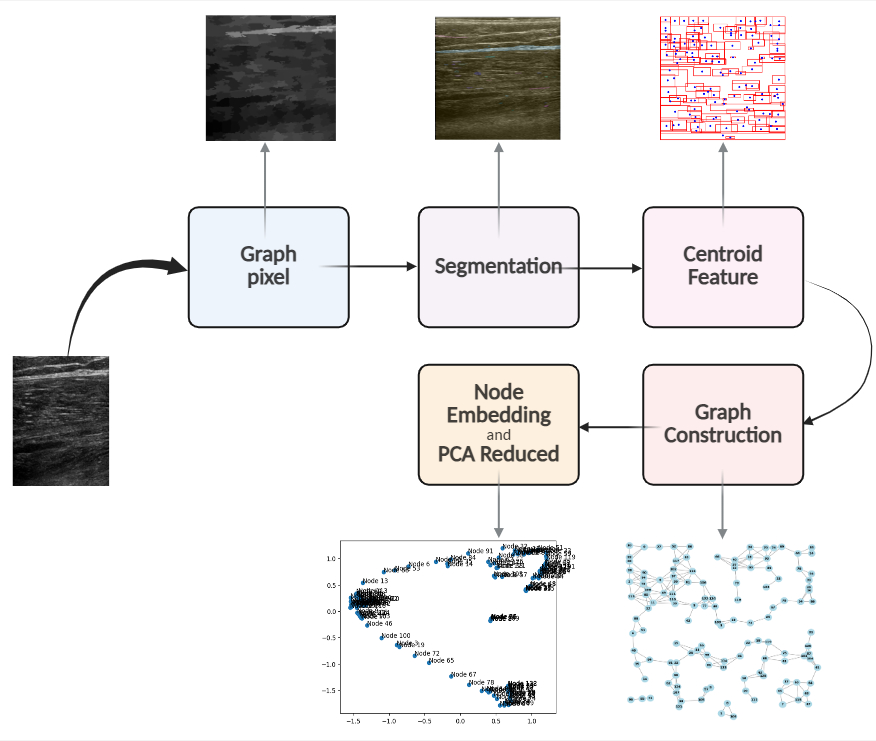}
    \caption{Superpixel-Based Graph Construction for Fine-Grained Analysis}
    \label{fig:enter-label}
\end{figure}
\subsubsection{\textbf{Numeric Feature Extraction } }
\label{sec2}
Structured clinical variables such as age, sex, height, weight, BMI, and SPPB score were first preprocessed to unify their format and scale. Categorical features were vectorized using appropriate encoding strategies (e.g., one-hot or trainable embeddings), while continuous ones were scaled to a normalized range to align with the representation space of other modalities. These harmonized vectors were then passed through several deep neural network variants, each producing a 400-dimensional latent embedding designed to capture both statistical and clinically relevant semantic patterns.
\subsubsection{\textbf{Textual Feature Extraction } }
\label{sec2}
To address the challenge of limited labeled data and sparse multimodal supervision in clinical scenarios particularly in sarcopenia diagnosis and management, we incorporate an external biomedical knowledge retrieval mechanism based on a customized RAG pipeline. This component enriches the model’s textual understanding by aligning patient-level numeric profiles with domain-specific medical evidence retrieved from scientific literature. The pipeline is designed with a two-phase mechanism: clinical-aware query construction and semantic filtering over retrieved content described below.
\begin{enumerate}
    \item {\textbf{Query Construction with Clinical-Aware Semantic Expansion } }
As explained earlier, each patient’s structured data point comprising clinical variables such as BMI, SPPB score, age, and gender is used to generate a contextualized query set. This is done by semantically expanding each variable into high-level biomedical concepts using the UMLS. For instance:
\begin{itemize}
    \item BMI up 30 is mapped to the concept Obesity or High Body Mass Index.
    \item SPPB down 7 reflects Low Physical Performance or Frailty.
    \item Age up 65 is expanded to Elderly Population, and Gender maps to Male/Female phenotypes.
\end{itemize}
Each variable-triggered concept is first mapped to its CUI via UMLS APIs. These CUIs are then incorporated into multiple PubMed search queries, leveraging structured templates that cover diverse clinical aspects including Diagnosis, Prognosis, Risk Factors, and Comorbidities. These enriched queries are then dispatched to PubMed to retrieve relevant abstracts in either real-time or batch mode. For each patient, a set of documents is returned, forming the initial knowledge corpus. As shown in Figure 4,  a semantic pipeline translates clinical inputs into PubMed queries, UMLS.
\begin{figure}
    \centering
    \scriptsize
    \includegraphics[width=1\linewidth]{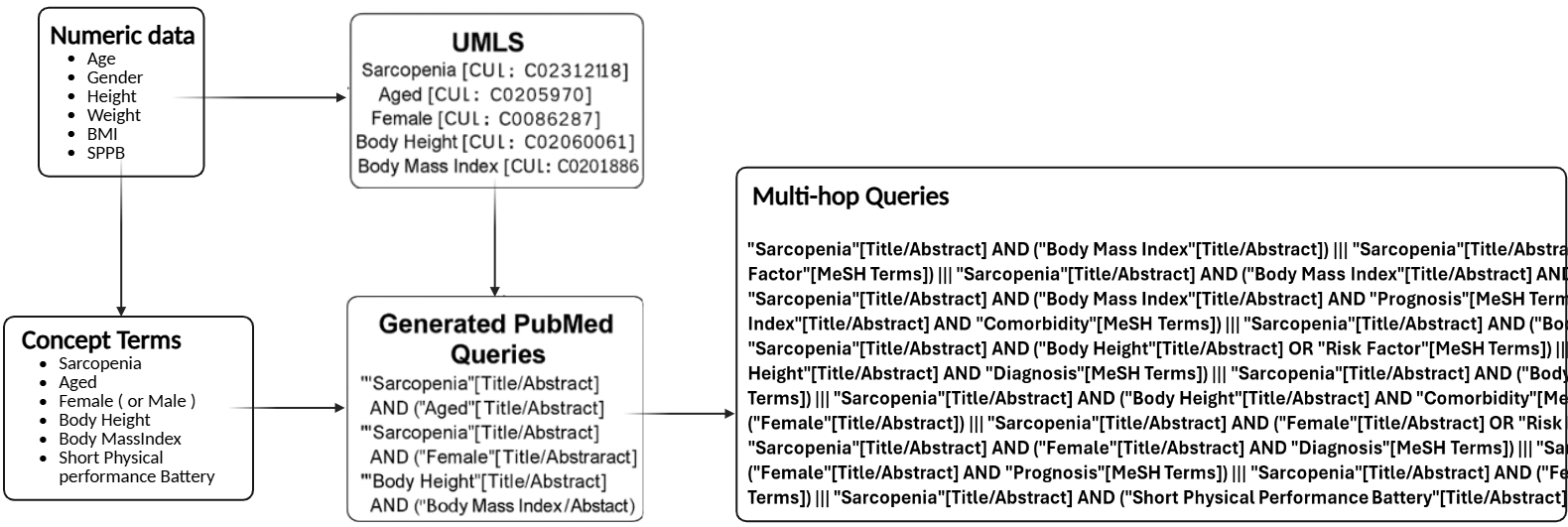}
    \caption{A semantic pipeline that translates clinical inputs into UMLS PubMed queries}
    \label{fig:enter-label}
\end{figure}
\item {\textbf{ Two-Stage Retrieval and Semantic Filtering Pipeline}}
To refine and prioritize the textual evidence retrieved in the previous stage from medical knowledge sources such as PubMed. We introduce a two-stage hierarchical filtering process that combines topic modeling and semantic similarity.
\begin{enumerate}
\item {\textbf{ Topic-Level Filtering via LDA:}}
The retrieved abstracts are tokenized into sentence-level units. Each sentence is passed through a LDA model, pre-trained on a large corpus of sarcopenia-related literature to identify major latent topics (for example: muscle degeneration, nutrition, mobility loss, and inflammatory markers). Each sentence is assigned a topic distribution vector, where T is the number of predefined topics. Simultaneously, the query is also mapped to a topic distribution using the same LDA model. Sentences with cosine similarity (or KL-divergence) above a certain threshold with respect to the query’s topic vector are retained. This step removes irrelevant content and ensures that only topically aligned information moves to the next stage.
\item {\textbf{ Semantic Filtering via Embedding-Based Similarity:}}
The topic-filtered sentences are embedded into high-dimensional vector space using pretrained domain-specific encoders such as PubMedBERT, BioBERT, BioClinicalBERT, BioELECTRA, BioLinkBERT, BiomedVLP, and MedCPT (w/ Article variant). These encoders are optimized for biomedical literature and allow fine-grained semantic comparisons. The current patient’s query is also encoded into the same vector space. Cosine similarity is computed between the query vector and all sentence vectors. The top-C semantically similar sentences (e.g., C = 10) are selected to serve as the final knowledge snippets. These top-C embeddings form a dense matrix of shape C × D ( We typically used D = 768, where D denotes the dimensionality of the text feature space ) representing the most relevant and personalized textual context for that patient. This matrix is used as a knowledge-enhanced textual feature and fused downstream with image and numeric modalities for decision-making or classification.
Example: For a 72-year-old female patient with BMI=32 and SPPB=5, the pipeline might retrieve sentences discussing obesity-related muscle loss in elderly women, mobility challenges, and predictive factors of frailty, all semantically aligned with the patient profile.
\end{enumerate}
To select appropriate biomedical language and vision-language models for our tasks, we evaluated various pretrained architectures. This comparison informed our choice of models based on dataset characteristics and clinical relevance. Table 1 summarizes the biomedical and vision-language models evaluated in this work. The models vary in their pretraining data, architecture, and intended use cases, enabling us to leverage complementary strengths for medical visual question answering and disease prediction on the VQA-RAD and Sarcopenia datasets.
\begin{table}[!t]
    \renewcommand{\arraystretch}{.9}
    \centering
    \scriptsize
    \caption{Biomedical and vision-language models with descriptions and selection rationale for VQA-RAD and Sarcopenia.}
    \resizebox{\textwidth}{!}{
    \begin{tabular}{|l|p{1.5cm}|p{8.2cm}|}
        \hline
        \textbf{Model} & \textbf{Description} & \textbf{Strength \& Rationale} \\ \hline
        BioClinicalBERT \cite{alsentzer2019publicly} & Fine-tuned on MIMIC-III clinical notes & The language used in VQA-RAD questions closely resembles clinical narratives, such as examination findings, patient conditions, or mentions of diseases and symptoms. BioClinicalBERT is well-suited to understand the structure and writing style of such medical notes. \\ \hline
        PubMedBERT \cite{zhang2025multimodal} & Pretrained from scratch on PubMed abstracts & Since many VQA-RAD questions are formulated in a scientific tone and include specialized medical terminology (e.g., atelectasis, cardiomegaly, consolidation), PubMedBERT provides the best lexical and semantic alignment. Its custom tokenizer allows for a highly precise understanding of scientific sentence structures. \\ \hline
        BioBERT \cite{deka2022evidence} & BERT fine-tuned on PubMed and PMC & Although not trained from scratch on biomedical texts like PubMedBERT, BioBERT shows strong medical language understanding compared to general-domain BERT. It performs well on questions that fall between general language and highly technical biomedical terms. \\ \hline
        BioLinkBERT \cite{yasunaga2022linkbert} & Integrates citation links in biomedical corpora & Ideal for multi-hop reasoning questions or those that require connecting dispersed concepts such as synthesizing different findings in the image to reach a diagnosis. Its structured pretraining helps it infer complex relations, such as combining “anomaly location” with “its impact on another structure.” \\ \hline
        BioELECTRA \cite{kanakarajan2021bioelectra} & Biomedical version of ELECTRA & Highly efficient and fast in extracting initial information from questions. Suitable for environments with limited computational resources or for preprocessing steps. \\ \hline
        MedCPT \cite{jin2023medcpt} & Instruction-tuned generative model & Offers strong capabilities in syntactic and interrogative understanding, along with generating fluent, well-structured answers. They are especially effective for descriptive or reasoning-based questions. \\ \hline
        MedCPT\_Article \cite{jin2023medcpt} & Full-text medical articles & Trained with full-length medical articles, allowing for better comprehension of long-form and context-rich questions. \\ \hline
        BiomedVLP \cite{bannur2023learning} & Jointly pretrained on image-text medical data & A vision-language model jointly trained on medical images and their associated textual descriptions. It enables cross-modal alignment between the question and the visual content, which is particularly useful when the question refers to a specific region in the image (e.g., “What is seen in the right upper lobe?”), thereby facilitating effective localization of relevant areas. \\ \hline
    \end{tabular} } 
    \label{tab:biomedical_models}
\end{table}
\item {\textbf{Fusion and Representation } }
\label{sec3}
To integrate diverse modalities of patient data with external biomedical knowledge, we construct a unified representation space. This fusion step aims to enhance the model's ability to reason over numeric features (e.g., age, BMI), imaging biomarkers, and textual evidence retrieved from biomedical literature. As shown in Figure 5, patient-specific queries are generated using demographic and clinical data, and relevant PubMed abstracts are retrieved and processed. These sentences are then semantically filtered using LDA and embedded via biomedical transformers to align with the patient’s clinical profile. The result is a dense matrix of shape C × D, where each row corresponds to an embedded sentence that semantically matches the patient query. This matrix is subsequently fused with the patient’s numeric and image-based features, grounding the model's predictions in trusted medical knowledge. This multimodal integration improves both interpretability and robustness, particularly in rare or ambiguous clinical cases.
\begin{figure}[!t]
    \centering
    \scriptsize
    \includegraphics[width=1\linewidth]{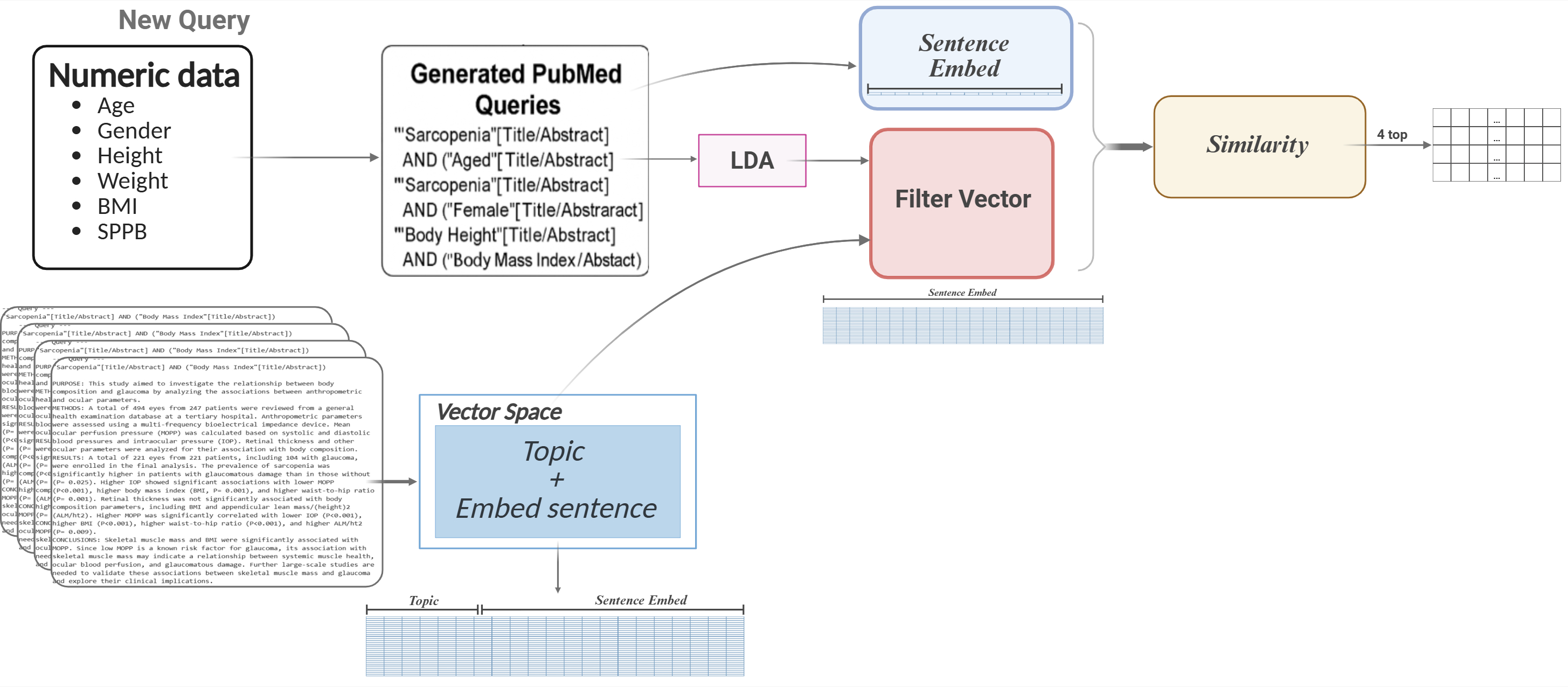}
    \caption{Semantic Filtering and Knowledge-Augmented Fusion}
    \label{fig:enter-label}
\end{figure}
\end{enumerate}
\subsection{\textbf{Model Development } }
\label{sec2}
We propose a dynamic feature selection framework that integrates hierarchical visual representations with textual features using a lightweight fusion mechanism inspired by LoRA and soft routing. The model is designed to adaptively select the most informative representation among coarse, mid-level, and fine-grained image features for each input instance based on the associated textual context.
To better explain the design of the model, it is expressed mathematically in the following. The overall pipeline consists of the following components:
\subsubsection{\textbf{Input and Feature Representation } }
\label{sec2}
Each input sample consists of three levels of image features:
\begin{align*}
\text{Coarse-level feature: }  \mathbf{v}^{(c)} &\in \mathbb{R}^{d_v} \quad \\
\text{Mid-level feature: }  \mathbf{v}^{(m)} &\in \mathbb{R}^{d_v} \quad \\
\text{Fine-grained feature: } \mathbf{v}^{(f)} &\in \mathbb{R}^{d_v} \quad 
\end{align*}
along with textual and numeric features:
\begin{align*}
\text{Textual feature: } \mathbf{t} &\in \mathbb{R}^{d_t} \quad \\
\text{Numeric feature: } \mathbf{n} &\in \mathbb{R}^{d_n} \quad 
\end{align*}
Here, $\mathbf{t}$ is extracted from the RAG-retrieved domain-specific medical text
and $\mathbf{n}$ is derived from structured numeric characteristics such as (e.g., age, BMI).

\subsubsection{\textbf{Text Projection } }
\label{sec2}
Text features are first projected into a common space with visual features:
\begin{equation}
\tilde{\mathbf{t}} = W_t \mathbf{t}+b_t , \quad \tilde{\mathbf{t}} \in \mathbb{R}^{d}
\end{equation}
where $\quad W_t \in \mathbb{R}^{d \times d_t}$ is the text projection weight,
$\tilde{\mathbf{t}}$ is the projected text feature, and
$d$ is the shared hidden dimension.
\subsubsection{\textbf{LoRA Module for Visual Features } }
\label{sec2}
Each visual feature is passed through a Low-Rank Adapter:
\begin{equation}
\mathrm{LoRA}(\mathbf{v}) = \mathbf{W}_{\text{up}} \left( \mathbf{W}_{\text{down}} \mathbf{v} + \mathbf{b}_{\text{down}} \right) + \mathbf{b}_{\text{up}}, 
\quad \mathbf{W}_{\text{up}} \in \mathbb{R}^{d \times r}, \quad \mathbf{W}_{\text{down}} \in \mathbb{R}^{r \times d}
\end{equation}
where,
 \begin{align*} 
 {d>>r} \text{ is the adapter rank,} \end{align*}  
 \begin{align*} {LoRA} \text{ learns a task-specific adaptation with a lightweight parameter footprint.} \quad 
 \end{align*}   
Applying LoRA on each visual feature:
\begin{equation}
\left\{
\begin{aligned}
\mathbf{h}^{(c)} &= \mathrm{LoRA}(\mathbf{v}^{(c)}) + \tilde{t} \\
\mathbf{h}^{(m)} &= \mathrm{LoRA}(\mathbf{v}^{(m)}) + \tilde{t} \\
\mathbf{h}^{(f)} &= \mathrm{LoRA}(\mathbf{v}^{(f)}) + \tilde{t}
\end{aligned}
\right.
\end{equation}

\subsubsection{\textbf{Gating Mechanism } }
\label{sec2}
A soft routing gate is computed from the projected text feature:
\begin{equation}
\mathbf{g} = \text{Softmax}(W_g \mathbf{\tilde{t}}+ b_g ) , \quad \mathbf{g} \in \mathbb{R}^3 
\end{equation}
Then the feature corresponding to the maximum gate value is selected:
\begin{equation}
\begin{aligned}
\mathbf{i^*} &= \arg\max(\mathbf{g})  , \quad 
\mathbf{h}_{\text{selected}} &=
\begin{cases}
\mathbf{h}^{(c)} & \text{if } \mathbf{i^*} = 0 \\
\mathbf{h}^{(m)} & \text{if } \mathbf{i^*} = 1 \\
\mathbf{h}^{(f)} & \text{if } \mathbf{i^*} = 2 \\
\end{cases}
\end{aligned}
\end{equation}
\subsubsection{\textbf{Final Representation and Classification } }
\label{sec2}
The final representation is computed as:
\begin{equation}
\mathbf{z} = \mathbf{h}_{\text{selected}} + \tilde{t}
\end{equation}
The classification is then performed as follows:
\begin{equation}
\mathbf {\hat{y}} = \text{Softmax}(W_c \mathbf{z}+ b_c)
\end{equation}
\subsubsection{\textbf{Loss Function } }
\label{sec2}
Training is done by minimizing the cross-entropy loss between predicted and true labels:
\begin{equation}
\mathcal{L} = - \sum_{i=1}^N y_i \log(\hat{y}_i)
\end{equation}

Given the limited size and domain-specific characteristics of the rare disease dataset, we intentionally opted against employing large-scale language models such as GPT-3 or LLaMA. (A comparative evaluation of these models is presented in the Results section.) Instead, we adopted a lightweight feature extraction approach using the T5 model. This decision is grounded in prior evidence suggesting that, in data-constrained settings like rare disease classification, smaller transformer-based architectures can offer distinct advantages. These include better generalization to unseen cases, improved interpretability of intermediate representations, and substantially lower computational demands. Furthermore, compact models reduce the likelihood of overfitting an especially important consideration when working with imbalanced or low-volume clinical datasets.
\subsection{\textbf{Model Evaluation } }
\label{sec2}
To assess the detection performance of the created classifiers, several metrics were employed, including the receiver operating characteristics curve (ROC), area under the receiver operating characteristics curve (AU-ROC), Precision, Recall, F1-Score, and detection accuracy. These performance metrics are determined by searching for the values of true positive (TP), false positive (FP), false negative (FN), and true negative (TN). We employed the following evaluation metrics across all datasets:

\textbf{ Accuracy:} Represents the proportion of correctly predicted answers out of the total number of samples. While useful for general evaluation, it may be insufficient when class distribution is imbalanced.
\begin{equation}
\text{Accuracy} = \frac{TP + TN}{TP + TN + FP + FN}
\end{equation}
\textbf{ F1-Score:} (Macro and Weighted): The macro-average F1-score treats all classes equally, providing insight into the model's ability to perform well across all answer types, regardless of frequency. The weighted F1-score, on the other hand, accounts for class imbalance by assigning weights proportional to class frequency, offering a more realistic evaluation in imbalanced settings.
\begin{equation}
\text{F1-score} = \frac{2 \cdot \text{Precision} \cdot \text{Recall}}{\text{Precision} + \text{Recall}}
\end{equation}

\textbf{ Precision and Recall:} These metrics are especially critical in the sarcopenia classification task due to the inherent class imbalance. High recall ensures that the model correctly identifies most of the positive (sarcopenia) cases, while precision reflects the proportion of true positives among all predicted positives both essential for clinical reliability.
\begin{equation}
\text{Precision} = \frac{TP}{TP + FP}, \qquad \\
\end{equation}

\begin{equation}
\text{Recall} = \frac{TP}{TP + FN}
\end{equation}
\textbf{ ROC-AUC:} (Receiver Operating Characteristic - Area Under Curve): This metric evaluates the model's ability to discriminate between classes across all possible classification thresholds, making it particularly valuable for binary tasks like sarcopenia detection where sensitivity and specificity must be jointly considered.
\section{\textbf{Results } }
\label{sec3}
In this section, we present a comprehensive evaluation of our proposed models and methods. We begin by describing the datasets used for experiments, followed by detailed model performance assessments. Comparative analyses with baseline methods are provided to highlight the improvements achieved. Finally, ablation studies are conducted to investigate the contributions of various model components, including hierarchical feature representations for both sarcopenia detection and the VQA-RAD visual question answering tasks.
\subsection{\textbf{Comparison with Baseline Methods } }
\label{sec3}
To evaluate the performance of the proposed MedVQA-TREE tree framework, we report classification accuracy across three benchmark datasets: two widely-used public MedVQA datasets (VQA-RAD and PathVQA) and sarcopenia dataset. The results presented in the following table highlight the model’s ability to generalize across diverse medical domains and question types. The high accuracy achieved on all datasets demonstrates the effectiveness of our hierarchical visual reasoning in producing clinically relevant answers. We compared our model with several baseline methods. As summarized in Table 2, our method outperformed all baselines. Our model also showed higher interpretability and localization accuracy, particularly in low-data settings. The baselines include recent and well-recognized multimodal models such as different versions of LLaVA-Med, GPT-4v, and BioMed CLIP-based approaches, which are commonly used in the medical imaging and VQA community. This selection enables us to benchmark our model against the current standards in the field. These baseline methods have reported their performance on widely-used datasets such as VQA-RAD and PathVQA, allowing for direct and fair comparison of results. The selected baselines cover a range of architectures and techniques, including deep learning models focused on image features and language-vision fusion methods. This variety ensures that the generalizability and robustness of our proposed method are evaluated across different types of approaches and conditions. Regarding comparison to previous works, several recent studies have reported results on similar datasets, which we incorporated in our comparative analysis. Notably, our model achieves significantly higher accuracy on the sarcopenia dataset, highlighting its effectiveness. Additionally, the sarcopenia dataset used in our experiments corresponds to the SAID (Sarcopenia AI Diagnosis) dataset.
\begin{table}[!t]
    \centering
    \scriptsize
    \caption{Performance comparison of classification accuracy between the proposed MedVQA-TREE model and baseline methods across three medical VQA datasets.}
    \label{tab:accuracy_comparison}
    \begin{tabular}{|l|l|c|}
        \hline
        \textbf{Dataset} & \textbf{Model} & \textbf{Accuracy} \\
        \hline
        \multirow{11}{*}{VQA-RAD} 
            & ELIXR & 0.69 \\
            & LLaVA-Med (From LLaVA) & 0.84 \\
            & LLaVA-Med (From Vicuna) & 0.81 \\
            & LLaVA-Med (BioMed CLIP) & 0.83 \\
            & MUMC & 0.84 \\
            & PeFoMed & 0.87 \\
            & LaPA & 0.86 \\
            & MPRgen\_PM & 0.80 \\
            & MPRdisc\_BAN & 0.80 \\
            & GPT-4v & 0.61 \\
            & \textbf{Our Model (MedVQA-TREE)} & \textbf{0.88} \\
        \hline
        \multirow{6}{*}{PathVQA} 
            & LLaVA-Med (From LLaVA) & 0.91 \\
            & LLaVA-Med (From Vicuna) & 0.91 \\
            & LLaVA-Med (BioMed CLIP) & 0.91 \\
            & MUMC & 0.90 \\
            & PeFoMed & 0.91 \\
            & \textbf{Our Model (MedVQA-TREE)} & \textbf{0.84} \\
        \hline
        \multirow{1}{*}{Our Dataset}
            & \textbf{Our Model (MedVQA-TREE)} & \textbf{0.99} \\
        \hline
    \end{tabular}
\end{table}
\subsection{\textbf{Ablation Study } }
\label{sec3}
To better understand the contribution of each component in our architecture, we conducted a detailed ablation study by progressively removing or replacing key modules. We first performed the analysis on our custom sarcopenia dataset to highlight the importance of each component in a rare disease setting. Subsequently, we validated the generalizability of the findings on the publicly available VQA-RAD dataset.
\subsubsection{\textbf{Hierarchical Component Performance Evaluation for Sarcopenia  } }
\label{sec3}
To understand the individual and combined contributions of each hierarchical level in our model, we conducted extensive k-fold cross-validation experiments. The results show how different levels ranging from global to fine-grained affect diagnostic performance, both independently and in fusion with clinical data. 
\begin{enumerate}
    \item \textbf{ Performance of Individual Hierarchical Levels (3-level)}
\textbf{ Level 1: Global High-Level Features:}
This level captures abstract global patterns, such as anatomical orientation (e.g., transverse vs. longitudinal). Although these features are relatively coarse, they still provide useful cues for differentiating cases with marked structural changes.Table 2 summarizes the classification performance of the first hierarchical level, which captures global high-level features, evaluated through k-fold cross-validation.
\begin{table}[!t]
    \centering
    \scriptsize
    \caption{Evaluation of classification performance at the global feature level (Level 1) using k-fold cross-validation.}
    \label{tab:level1_kfold}
    \begin{tabular}{|l|c|c|c|c|c|}
        \hline
        \textbf{k-fold} & \textbf{Accuracy} & \textbf{Precision} & \textbf{Recall} & \textbf{F1-score} & \textbf{ROC AUC} \\
        \hline
        1-fold & 0.66 & 0.64 & 0.66 & 0.64 & 0.66 \\
        2-fold & 0.72 & 0.69 & 0.70 & 0.69 & 0.70 \\
        3-fold & 0.68 & 0.64 & 0.66 & 0.65 & 0.66 \\
        4-fold & 0.73 & 0.70 & 0.71 & 0.71 & 0.71 \\
        5-fold & 0.72 & 0.67 & 0.71 & 0.68 & 0.71 \\
        \textbf{Average} & \textbf{0.70} & \textbf{0.66} & \textbf{0.68} & \textbf{0.67} & \textbf{0.69} \\
        \hline
    \end{tabular}
\end{table}

\textbf{ Level 2: Regional Mid-Level Features:}
Using SAM-based segmentation, we extracted features from muscle regions. This level enhanced the model’s ability to detect localized muscle degradation. Table 4 reports the classification metrics obtained at the second hierarchical level, where mid-level regional features were extracted from segmented muscle regions using a SAM-based approach. These features provide localized insights into muscle condition, especially for detecting regional degradation patterns indicative of sarcopenia.

\begin{figure}[!t]
    \centering
    \scriptsize
    \includegraphics[width=0.6\linewidth]{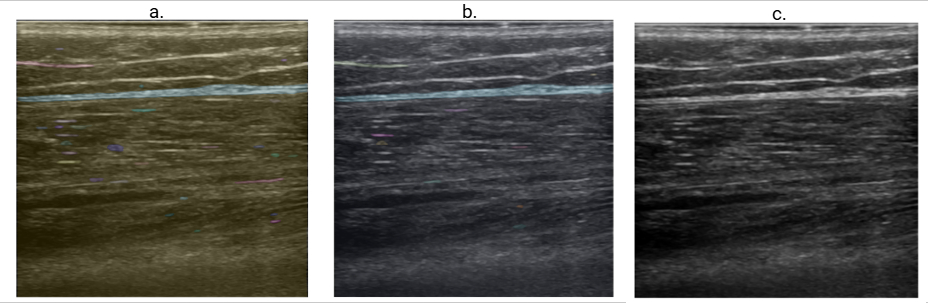}
    \caption{Comparison of segmentation results using different SAM-based models: (a) original SAM, (b) Med-SAM \cite{zhang2024segment}, and (c) F-SAM \cite{kirillov2023segment}}
    \label{fig:enter-label}
\end{figure}
\begin{figure}[!t]
    \centering
    \scriptsize
    \includegraphics[width=0.6\linewidth]{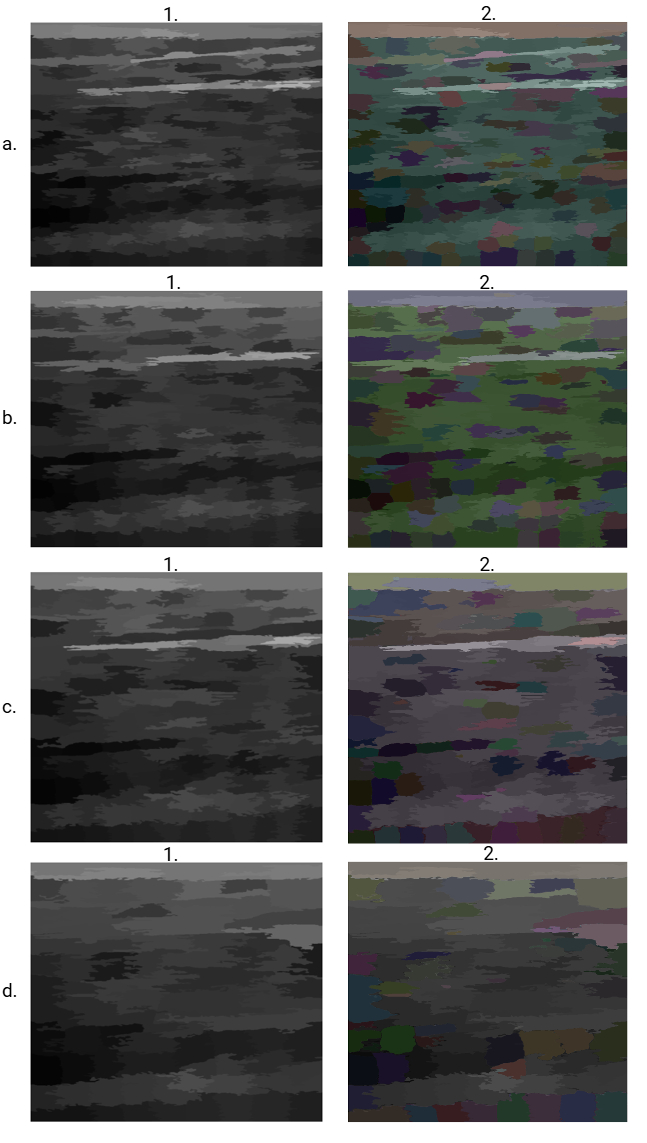}
    \caption{Segmentation comparison between the constructed superpixel graph and the SAM model outputs at varying levels of granularity. Subfigures illustrate the effect of segment count on the quality and structure of segmentation: (a) 250 segments, (b) 200 segments, (c) 150 segments, and (d) 100 segments. For each case, the left side presents the superpixel-based graph structure, while the right side shows the corresponding SAM-based segmentation result. The comparison highlights how finer or coarser superpixel division influences regional delineation and graph definition.}
    \label{fig:enter-label}
\end{figure}
\begin{table}[!t]
    \centering
    \scriptsize
    \caption{Classification performance of Level 2 (Regional Mid-Level Features) based on SAM-based muscle segmentation across k-fold cross-validation.}
    \label{tab:level2_kfold}
    \begin{tabular}{|l|c|c|c|c|c|}
        \hline
        \textbf{k-fold} & \textbf{Accuracy} & \textbf{Precision} & \textbf{Recall} & \textbf{F1-score} & \textbf{ROC AUC} \\
        \hline
        1-fold & 0.51 & 0.52 & 0.53 & 0.50 & 0.53 \\
        2-fold & 0.52 & 0.53 & 0.53 & 0.51 & 0.53 \\
        3-fold & 0.52 & 0.52 & 0.53 & 0.50 & 0.53 \\
        4-fold & 0.52 & 0.53 & 0.53 & 0.51 & 0.53 \\
        5-fold & 0.55 & 0.53 & 0.54 & 0.52 & 0.54 \\
        \textbf{Average} & \textbf{0.52} & \textbf{0.52} & \textbf{0.53} & \textbf{0.50} & \textbf{0.53} \\
        \hline
    \end{tabular}
\end{table}

\textbf{ Level 3 – Fine-Grained Spatial Graph Features:}

At the most detailed level, we employed superpixel segmentation combined with centroid-based spatial graph modeling to capture fine-grained inter-regional dependencies among anatomical components in the image. The process involved segmenting the image into coherent superpixels, extracting centroids from the masked regions, and constructing a spatial graph whose nodes represented anatomical segments, while the edges modeled their spatial relationships. Node embeddings were subsequently generated and converted into fixed-length vectors to be used in downstream classification tasks.
This level proved especially effective in borderline or ambiguous cases, where subtle signs of degeneration or localized structural abnormalities were present but not prominent in coarse-level analysis. Unlike coarser levels, this fine-grained modeling was capable of capturing subtle spatial patterns that might indicate early-stage sarcopenia or muscle degradation, contributing to a more nuanced representation of muscular health. 
Table 5 presents the classification results for the third and most detailed level in our hierarchical framework. At this stage, we used fine-grained spatial graph features derived from superpixel segmentation and centroid-based graph construction to capture inter-regional anatomical dependencies. This modeling allowed the system to detect nuanced spatial patterns, offering significant benefits in early or ambiguous cases of muscle degeneration.
\begin{table}[!t]
    \centering
    \scriptsize
    \caption{Performance of Level 3 (Fine-Grained Spatial Graph Features) using superpixel-based spatial graph modeling across k-folds.}
    \label{tab:level3_kfold}
    \begin{tabular}{|l|c|c|c|c|c|}
        \hline
        \textbf{k-fold} & \textbf{Accuracy} & \textbf{Precision} & \textbf{Recall} & \textbf{F1-score} & \textbf{ROC AUC} \\
        \hline
        1-fold & 0.68 & 0.34 & 0.50 & 0.40 & 0.50 \\
        2-fold & 0.68 & 0.34 & 0.50 & 0.40 & 0.50 \\
        3-fold & 0.69 & 0.84 & 0.50 & 0.41 & 0.50 \\
        4-fold & 0.67 & 0.58 & 0.50 & 0.40 & 0.50 \\
        5-fold & 0.73 & 0.36 & 0.49 & 0.42 & 0.49 \\
        \textbf{Average} & \textbf{0.68} & \textbf{0.49} & \textbf{0.50} & \textbf{0.40} & \textbf{0.50} \\
        \hline
    \end{tabular}
\end{table}
Overall, Level 3 modeling provided a granular and spatially-aware feature representation, which proved valuable in situations where broader-level cues were insufficient. This highlights the potential of graph-based anatomical modeling in enhancing the sensitivity of diagnostic models to localized and subtle pathological cues.
\item \textbf{ Impact of Combining Hierarchical Levels: }
To assess the benefit of hierarchical integration, we combined features extracted from all three visual levels: coarse, mid-level, and fine-grained spatial representations. This multi-level fusion enabled the model to utilize both high-level contextual cues and fine-grained local dependencies.
To evaluate the impact of hierarchical feature integration, we fused representations from all three visual abstraction levels coarse (global), mid-level (regional), and fine-grained (graph-based). This unified representation allowed the model to leverage both broad contextual understanding and localized spatial details. As shown in Table 6, this multi-level fusion led to a significant improvement across all evaluation metrics. Notably, the average accuracy reached 0.90, and the F1-score and ROC AUC both improved to 0.88, outperforming individual levels. These results demonstrate the complementary nature of hierarchical features and highlight their importance for robust performance, especially in complex or borderline cases.
\begin{table}[!t]
    \centering
    \scriptsize
    \caption{Performance of Combined Multi-Level Visual Features (Hierarchical Fusion) across k-folds.}
    \label{tab:level_combined}
    \begin{tabular}{|l|c|c|c|c|c|}
        \hline
        \textbf{k-fold} & \textbf{Accuracy} & \textbf{Precision} & \textbf{Recall} & \textbf{F1-score} & \textbf{ROC AUC} \\
        \hline
        1-fold & 0.89 & 0.88 & 0.86 & 0.87 & 0.86 \\
        2-fold & 0.91 & 0.89 & 0.90 & 0.90 & 0.90 \\
        3-fold & 0.90 & 0.89 & 0.88 & 0.89 & 0.88 \\
        4-fold & 0.90 & 0.88 & 0.89 & 0.89 & 0.89 \\
        5-fold & 0.90 & 0.88 & 0.87 & 0.88 & 0.87 \\
        \textbf{Average} & \textbf{0.90} & \textbf{0.88} & \textbf{0.88} & \textbf{0.88} & \textbf{0.88} \\
        \hline
    \end{tabular}
\end{table}

\item \textbf{ Role of Numerical Clinical Data:}
In the next step, we explored whether incorporating structured numerical clinical data (e.g., age, BMI, comorbidities) could enhance classification performance when combined with image-based features. Each level was tested independently with the clinical data.

\textbf{ Level 1 + Numerical Clinical Data:} Modest improvement, indicating that low-resolution features benefit from additional context.
\begin{table}[!t]
    \centering
    \scriptsize
    \caption{Performance of Level 1 Visual Features Combined with Numerical Clinical Data across k-folds.}
    \label{tab:level1_numerical}
    \begin{tabular}{|l|c|c|c|c|c|}
        \hline
        \textbf{k-fold} & \textbf{Accuracy} & \textbf{Precision} & \textbf{Recall} & \textbf{F1-score} & \textbf{ROC AUC} \\
        \hline
        1-fold & 0.65 & 0.63 & 0.64 & 0.63 & 0.64 \\
        2-fold & 0.70 & 0.66 & 0.68 & 0.67 & 0.68 \\
        3-fold & 0.68 & 0.65 & 0.67 & 0.65 & 0.67 \\
        4-fold & 0.73 & 0.70 & 0.71 & 0.70 & 0.71 \\
        5-fold & 0.73 & 0.68 & 0.71 & 0.69 & 0.71 \\
        \textbf{Average} & \textbf{0.69} & \textbf{0.66} & \textbf{0.68} & \textbf{0.66} & \textbf{0.68} \\
        \hline
    \end{tabular}
\end{table}

The results shown in Table 7, reveal a modest but consistent improvement in classification performance compared to using Level 1 features alone (see Table 3). This suggests that global visual patterns, although coarse, can be effectively contextualized using patient-specific structured data, particularly in cases where visual cues alone may be insufficient.

\textbf{ Level 2 + Numerical Clinical Data:} Performance remained moderate, suggesting that mid-level features alone may lack sufficient discriminative power.
\begin{table}[!t]
    \centering
    \scriptsize
    \caption{Performance of Level 2 Visual Features Combined with Numerical Clinical Data across k-folds.}
    \label{tab:level2_numerical}
    \begin{tabular}{|l|c|c|c|c|c|}
        \hline
        \textbf{k-fold} & \textbf{Accuracy} & \textbf{Precision} & \textbf{Recall} & \textbf{F1-score} & \textbf{ROC AUC} \\
        \hline
        1-fold & 0.59 & 0.57 & 0.58 & 0.57 & 0.58 \\
        2-fold & 0.59 & 0.57 & 0.59 & 0.57 & 0.59 \\
        3-fold & 0.62 & 0.59 & 0.60 & 0.59 & 0.60 \\
        4-fold & 0.62 & 0.61 & 0.62 & 0.60 & 0.62 \\
        5-fold & 0.60 & 0.57 & 0.59 & 0.56 & 0.59 \\
        \textbf{Average} & \textbf{0.60} & \textbf{0.58} & \textbf{0.60} & \textbf{0.58} & \textbf{0.60} \\
        \hline
    \end{tabular}
\end{table}
\textbf{ Level 3 + Numerical Clinical Data:} Significant gains, especially in borderline cases, reflecting the complementarity between detailed spatial features and clinical indicators.
\begin{table}[!t]
    \centering
    \scriptsize
    \caption{Performance of Level 3 Visual Features Combined with Numerical Clinical Data across k-folds.}
    \label{tab:level3_numerical}
    \begin{tabular}{|l|c|c|c|c|c|}
        \hline
        \textbf{k-fold} & \textbf{Accuracy} & \textbf{Precision} & \textbf{Recall} & \textbf{F1-score} & \textbf{ROC AUC} \\
        \hline
        1-fold & 0.87 & 0.85 & 0.85 & 0.85 & 0.85 \\
        2-fold & 0.89 & 0.88 & 0.87 & 0.88 & 0.87 \\
        3-fold & 0.90 & 0.88 & 0.89 & 0.88 & 0.89 \\
        4-fold & 0.86 & 0.85 & 0.85 & 0.85 & 0.85 \\
        5-fold & 0.88 & 0.85 & 0.85 & 0.85 & 0.85 \\
        \textbf{Average} & \textbf{0.88} & \textbf{0.86} & \textbf{0.86} & \textbf{0.86} & \textbf{0.86} \\
        \hline
    \end{tabular}
\end{table}
These results indicate that numerical clinical data help the model adaptively modulate its attention, allowing it to rely on coarser cues when sufficient and deeper representations when necessary thus improving interpretability and efficiency. 
\item \textbf{ Fusion Strategies without RAG } 
To further integrate multimodal information combining features from all visual levels with numerical clinical vectors. This fusion was applied at the embedding level before classification. The results demonstrated a notable improvement across all folds, achieving an average accuracy of 0.93 and ROC AUC of 0.93. This highlights the power of combining complementary information sources in a unified representation, allowing the model to make more confident and generalizable predictions.
\begin{table}[!t]
    \centering
    \scriptsize
    \caption{Performance of Embedding-Level Fusion of Multi-Level Visual Features and Clinical Data (Without RAG) across k-folds.}
    \label{tab:fusion_without_rag}
    \begin{tabular}{|l|c|c|c|c|c|}
        \hline
        \textbf{k-fold} & \textbf{Accuracy} & \textbf{Precision} & \textbf{Recall} & \textbf{F1-score} & \textbf{ROC AUC} \\
        \hline
        1-fold & 0.93 & 0.93 & 0.92 & 0.92 & 0.92 \\
        2-fold & 0.93 & 0.92 & 0.92 & 0.92 & 0.92 \\
        3-fold & 0.94 & 0.93 & 0.94 & 0.93 & 0.94 \\
        4-fold & 0.93 & 0.92 & 0.93 & 0.93 & 0.93 \\
        5-fold & 0.94 & 0.93 & 0.93 & 0.93 & 0.93 \\
        \textbf{Average} & \textbf{0.93} & \textbf{0.93} & \textbf{0.93} & \textbf{0.93} & \textbf{0.93} \\
        \hline
    \end{tabular}
\end{table}
Despite the strong performance achieved through direct fusion of visual and numerical features (average accuracy and ROC AUC of 0.93), this approach remains limited by its reliance solely on encoded input data. It lacks the capacity to adaptively incorporate external domain knowledge or context-specific cues. In contrast, RAG-based strategies allow dynamic retrieval of relevant biomedical knowledge, enabling the model to enhance its reasoning in ambiguous or complex diagnostic cases.
\item \textbf{ Fusion Strategies with RAG}
To overcome the limitations of static fusion approaches, we incorporated a RAG-based strategy. During training, clinically relevant queries were generated from the training data using medical knowledge bases such as UMLS. These queries were expanded across key dimensions including diagnostic criteria, comorbidities, risk factors, and disease progression and chunked at the sentence level. Using LDA topic modeling, these chunks were categorized and aligned with the training set. For test-time inference, a simplified query was generated from the patient record. This query was matched with relevant chunks in the training-aligned categories using topic-guided similarity. The resulting similarity matrix was concatenated with visual features (just one level by using gate selection) and numerical clinical data, and passed through a lightweight LoRA-enhanced network for classification.
This adaptive, knowledge-guided mechanism significantly improved the model’s contextual understanding and diagnostic precision. As shown in the table below, the model achieved near-perfect performance across all metrics, with an average accuracy, precision, recall, F1-score, and ROC AUC of 0.99.
\begin{table}[!t]
    \centering
    \scriptsize
    \caption{Performance of Embedding-Level Fusion of Multi-Level Visual Features and Clinical Data (With RAG).}
    \label{tab:fusion_with_rag}
    \begin{tabular}{|l|c|c|c|c|c|}
        \hline
        \textbf{Method} & \textbf{Accuracy} & \textbf{Precision} & \textbf{Recall} & \textbf{F1-score} & \textbf{ROC AUC} \\
        \hline
        \textbf{Average (With RAG)} & \textbf{0.99} & \textbf{0.99} & \textbf{0.99} & \textbf{0.99} & \textbf{0.99} \\
        \hline
    \end{tabular}
\end{table}
In the context of medical text processing, resources such as UMLS and WordNet are commonly used to extract synonyms, abbreviations, and the semantic meaning of terms. Incorporating semantic concepts enables the model to recognize similarities between sentences that use different surface words but convey the same meaning. By enriching the input with these concepts and retrieving formal definitions from medical knowledge bases using large language models, we enhance semantic understanding and representation learning. Additionally, instead of relying on meaningless index-based retrieval, we grouped the retrieved documents based on topics extracted using LDA, which significantly improved retrieval precision, reduced computational overhead, and accelerated inference speed. This topic-guided retrieval made the knowledge integration process more focused and efficient.  In order to further investigate the generalizability of our image feature extraction model, we performed ablation studies on the broader VQA-RAD dataset.
\end{enumerate}

\subsubsection{\textbf{Hierarchical Component Performance Evaluation for VQA-RAD } }
\label{sec3}
To evaluate the individual contributions of different levels of visual representation, we conducted ablation studies across five folds, removing the LoRA fusion mechanism and gating module in each configuration. The results reveal that using only the pretrained backbone without hierarchical decomposition performs moderately (average accuracy 0.73), but precision and recall remain low, reflecting limited task-specific adaptation. The coarse-level features (entire image) yield the highest accuracy among individual levels (up to 0.75), indicating that global contextual cues are informative; however, low F1-scores confirm poor sensitivity and class balance. The mid-level features (regional/segmented areas) also show decent accuracy but still struggle with generalization due to insufficient fine-scale integration. In contrast, the fine-grained spatial graph alone performs poorly across all metrics (F1-score 0.10–0.13), underscoring that local features lack sufficient discriminative power without broader context. These findings highlight the complementary nature of hierarchical features and justify the need for feature fusion via LoRA and gating mechanisms, as later shown to drastically improve performance.
\begin{table}[!t]
    \centering
    \scriptsize
    \caption{Performance of Only Pretrained Model (Without Gate Selection and LoRA).}
    \label{tab:pretrained_no_gate_lora}
    \begin{tabular}{|l|c|c|c|c|}
        \hline
        \textbf{K-Fold} & \textbf{Accuracy} & \textbf{Precision} & \textbf{Recall} & \textbf{F1-score} \\
        \hline
        1-Fold & 0.70 & 0.16 & 0.19 & 0.17 \\
        2-Fold & 0.74 & 0.16 & 0.18 & 0.16 \\
        3-Fold & 0.75 & 0.12 & 0.12 & 0.12 \\
        4-Fold & 0.70 & 0.14 & 0.14 & 0.14 \\
        5-Fold & 0.72 & 0.17 & 0.17 & 0.16 \\
        \hline
    \end{tabular}
\end{table}
\begin{table}[!t]
    \centering
    \scriptsize
    \caption{Performance of Only Coarse-Level Visual Features (Without Gate Selection and LoRA).}
    \label{tab:coarse_no_gate_lora}
    \begin{tabular}{|l|c|c|c|c|}
        \hline
        \textbf{K-Fold} & \textbf{Accuracy} & \textbf{Precision} & \textbf{Recall} & \textbf{F1-score} \\
        \hline
        1-Fold & 0.71 & 0.12 & 0.15 & 0.12 \\
        2-Fold & 0.75 & 0.15 & 0.17 & 0.16 \\
        3-Fold & 0.75 & 0.14 & 0.17 & 0.15 \\
        4-Fold & 0.70 & 0.14 & 0.15 & 0.15 \\
        5-Fold & 0.72 & 0.19 & 0.19 & 0.18 \\
        \hline
    \end{tabular}
\end{table}
\begin{table}[!t]
    \centering
    \scriptsize
    \caption{Performance of Only Mid-Level Visual Features (Without Gate Selection and LoRA).}
    \label{tab:mid_no_gate_lora}
    \begin{tabular}{|l|c|c|c|c|}
        \hline
        \textbf{K-Fold} & \textbf{Accuracy} & \textbf{Precision} & \textbf{Recall} & \textbf{F1-score} \\
        \hline
        1-Fold & 0.70 & 0.12 & 0.15 & 0.12 \\
        2-Fold & 0.73 & 0.16 & 0.18 & 0.16 \\
        3-Fold & 0.73 & 0.13 & 0.13 & 0.13 \\
        4-Fold & 0.70 & 0.13 & 0.13 & 0.13 \\
        5-Fold & 0.71 & 0.13 & 0.12 & 0.12 \\
        \hline
    \end{tabular}
\end{table}
\begin{table}[!t]
    \centering
    \scriptsize
    \caption{Performance of Only Fine-Grained Visual Features (Without Gate Selection and LoRA).}
    \label{tab:fine_no_gate_lora}
    \begin{tabular}{|l|c|c|c|c|}
        \hline
        \textbf{K-Fold} & \textbf{Accuracy} & \textbf{Precision} & \textbf{Recall} & \textbf{F1-score} \\
        \hline
        1-Fold & 0.62 & 0.05 & 0.06 & 0.05 \\
        2-Fold & 0.65 & 0.10 & 0.10 & 0.10 \\
        3-Fold & 0.67 & 0.13 & 0.13 & 0.13 \\
        4-Fold & 0.64 & 0.10 & 0.11 & 0.11 \\
        5-Fold & 0.63 & 0.08 & 0.09 & 0.08 \\
        \hline
    \end{tabular}
\end{table}

\textbf{ Rationale for Selecting Biomedical Language Models for VQA-RAD}

To assess the effectiveness of biomedical language models in VQA tasks, we evaluated a diverse set of domain-specific models on the VQA-RAD dataset. This dataset comprises complex and fine-grained medical visual questions that demand both domain expertise and multi-modal reasoning capabilities. General-purpose language models often fall short in understanding radiological terminology, anatomical references, and pathology-specific descriptions. Hence, we selected models that are pretrained or fine-tuned on biomedical corpora, clinical texts, or vision-language data. We included encoder-only models (e.g., BioClinicalBERT, PubMedBERT, BioBERT, BioLinkBERT, BioELECTRA), decoder-only models (e.g., MedCPT, MedCPT-Article), and multimodal encoder-decoder models (e.g., BiomedVLP) to provide a comprehensive architectural comparison and determine which paradigm best supports hierarchical visual reasoning in medical VQA.

\begin{table}[!t]
    \centering
    \scriptsize
    \caption{Evaluation Metrics and Performance Gaps of Biomedical Language Models for Visual Question Answering on VQA-RAD.}
    \label{tab:biomedical_vqa_models}
    \begin{tabular}{|l|c|c|c|c|c|c|}
        \hline
        \textbf{Model} & \textbf{Accuracy} & \textbf{Precision} & \textbf{Recall} & \textbf{F1-score} & \textbf{Accuracy Gap} & \textbf{F1 Gap} \\
        \hline
        BioClinicalBERT     & 0.70 & 0.77 & 0.79 & 0.77 & 0.01 & 0.07 \\
        PubMedBERT          & 0.81 & 0.79 & 0.80 & 0.78 & 0.01 & 0.06 \\
        BioBERT             & 0.68 & 0.72 & 0.73 & 0.71 & 0.02 & 0.07 \\
        BiomedVLP           & 0.70 & 0.54 & 0.55 & 0.54 & 0.01 & 0.21 \\
        MedCPT              & 0.78 & 0.72 & 0.73 & 0.72 & 0.02 & 0.12 \\
        MedCPT\_Article     & 0.72 & 0.66 & 0.68 & 0.66 & 0.02 & 0.15 \\
        BioELECTRA          & 0.65 & 0.64 & 0.64 & 0.63 & 0.03 & 0.16 \\
        BioLinkBERT         & 0.74 & 0.84 & 0.84 & 0.83 & 0.01 & 0.07 \\
        \hline
    \end{tabular}
\end{table}
Based on the evaluation metrics, PubMedBERT achieved the highest overall accuracy (0.8187) along with a strong F1-score (0.7820), indicating robust performance across metrics. Meanwhile, BioLinkBERT attained the highest F1-score (0.8395), supported by excellent precision and recall, although its slightly lower accuracy might be due to class imbalance. Models like BiomedVLP, BioELECTRA, and MedCPT-Article showed significant gaps between training and validation F1-scores for example, BiomedVLP exhibited a gap of 0.2101 suggesting potential overfitting. These models may benefit from additional regularization techniques or early stopping strategies to improve generalization. In contrast, PubMedBERT, BioClinicalBERT, and BioLinkBERT demonstrated more stable and generalizable performance, as reflected by their minimal train-validation performance gaps. On the lower end, BioELECTRA and BiomedVLP exhibited the weakest results, with the lowest accuracy and F1-scores overall. Furthermore, applying LoRA in the multi-modal fusion framework improved accuracy from 0.8266 (without LoRA) to 0.8887, underscoring the effectiveness of this fusion technique in enhancing model performance.
\begin{table}[!t]
    \centering
    \scriptsize
    \caption{Impact of LoRA Fusion on Model Accuracy in Multi-Modal VQA}
    \label{tab:lora_impact}
    \begin{tabular}{|l|c|}
        \hline
        \textbf{Model - Metric} & \textbf{Accuracy} \\
        \hline
        With LoRA in Multi-Modal Fusion    & 0.88 \\
        Without LoRA in Multi-Modal Fusion & 0.82 \\
        \hline
    \end{tabular}
\end{table}
\section{\textbf{Discussion } }
\label{sec4}
This section interprets and contextualizes the key findings of the study. We first summarize the main results, emphasizing their significance and implications for the field. Next, we compare our outcomes with those reported in previous studies, highlighting similarities, differences, and potential reasons for discrepancies. Finally, we discuss the limitations of our current work and propose directions for future research to address these challenges and further improve performance.
\subsection{\textbf{Summary of Findings } }
\label{sec4}
This section outlines the main design components of our system, aimed at providing an interpretable, computationally efficient, and practically deployable approach for sarcopenia diagnosis. Rather than emphasizing scale or complexity, our focus is on adaptability, integration of multimodal data, and alignment with clinical needs.
\subsubsection{\textbf{Adaptive Hierarchical Feature Extraction } }
\label{sec4}
The framework dynamically adjusts the depth of image processing based on the informativeness of the available clinical data. Unlike models that apply a uniform level of processing across all modalities, our approach selectively scales visual feature extraction depending on the clinical context. This can help prioritize computational resources for more complex or ambiguous cases and reduce unnecessary processing when clinical indicators are already informative.
\subsubsection{\textbf{Zero-Shot Use of Segmentation Models (SAM / MedSAM) } }
\label{sec4}
We utilized general-purpose segmentation models such as SAM and MedSAM in a zero-shot manner, without task-specific retraining. This choice helps simplify deployment and reduces dependency on annotated training data, particularly in scenarios where data collection is constrained. It also aligns with recent trends in medical imaging, where pre-trained foundation models are increasingly used for flexible segmentation tasks.
\subsubsection{\textbf{Anatomically-Informed Graph Construction for Spatial Reasoning } }
\label{sec4}
Following segmentation of relevant muscle regions, we constructed spatial graphs to model anatomical relationships such as symmetry, alignment, and relative positioning. This structure supports the learning of clinically relevant spatial patterns and adds a level of interpretability not easily captured by conventional CNN-based approaches. The use of graph reasoning aims to support more informed predictions in sarcopenia assessment.
\subsubsection{\textbf{Knowledge-Augmented Reasoning via Semantic and Multi-Hop Retrieval } }
\label{sec4}
To address complex medical queries, we developed a knowledge retrieval module based on UMLS-guided multi-query generation. It combines sentence-level semantic embeddings with topic-based LDA filtering to retrieve relevant external information efficiently. Unlike earlier VQA systems that often rely on shallow retrieval or static knowledge, our method supports multi-hop reasoning and dynamic access to external content in a more controlled and explainable manner.
\subsubsection{\textbf{Robustness in Low-Resource Clinical Settings } }
\label{sec4}
Instead of relying on large-scale models like GPT-3 or LLaMA, we selected a smaller model (T5-small/base) and fine-tuned it using LoRA. This approach reflects a focus on practical deployment in real-world clinical environments where data and computational resources are often limited. Despite its lightweight design, the model demonstrates acceptable performance under low-resource conditions, suggesting potential for broader applicability in similar clinical contexts.
\subsection{\textbf{Comparison to previous study results } }
\label{sec4}
Given the size and domain-specific nature of our dataset, we opted for smaller models such as T5-small/base rather than large-scale language models like GPT-3 or LLaMA. Previous work has shown that compact transformer-based models can perform competitively on medical classification tasks, particularly when annotated data is limited. A recent scoping review in healthcare NLP reports that domain-adapted smaller models can match or exceed the performance of larger ones in terms of accuracy, while offering benefits such as interpretability and lower inference cost. For example, studies using ClinicalT5 variants indicate that ClinicalT5-base (~220M parameters) achieves adequate performance in tasks like NER, classification, and risk prediction, without the overhead associated with larger models. In addition, smaller models are less prone to overfitting and are more compatible with clinical environments that have constrained computing resources. Our model selection was based on these practical considerations, with an emphasis on interpretability and resource efficiency rather than scale \cite{rohanian2024lightweight, lu2022clinicalt5, ao2025comparative}.

When compared to recent VQA systems such as ACMA and OphGLM, our model employs a different design focus. ACMA uses asymmetric cross-modal attention but does not incorporate hierarchical visual processing or external clinical knowledge. OphGLM is tailored for ophthalmology and operates primarily with static image-text pairs, lacking dynamic retrieval. Our system introduces a level of adaptability modifying both the granularity of image analysis and the depth of knowledge retrieval based on the input question which may be beneficial for multimodal tasks like sarcopenia diagnosis \cite{li2023asymmetric, deng2024ophglm}.

Recent research on multi-modal, multi-granularity fusion has shown promising results in other areas, such as brain network analysis for epilepsy detection using MMF-NNs. These models combine structural and functional MRI features at edge, node, and graph levels to model topology more effectively. While that work targets brain networks, the principle of multi-level fusion is relevant to our study. Our model applies a similar fusion strategy, but across heterogeneous data types visual features from different levels of image processing (global features, segmentations, spatial graphs) and clinical numerical data. Furthermore, MMF-NNs emphasize interactive feature learning at lower topological levels, while our method introduces dynamic gating mechanisms that select the most relevant feature level for the given question. This allows the model to adjust computation based on task complexity and clinical context, potentially improving efficiency and interpretability in resource-constrained settings \cite{huang2024mmf}.
\subsubsection{\textbf{Visual and Clinical Data Fusion } }
\label{sec4}
Many earlier approaches used general-purpose models (e.g., VGG, ResNet, standard language models) to extract features separately from image and text inputs. Our method introduces an adaptive framework that adjusts image processing depth based on the informativeness of accompanying clinical data. It extracts features at three levels: global orientation, anatomical segmentation, and spatial graph representations. This layered approach allows the model to reduce or expand visual processing depending on the available context.

Compared to prior systems such as CPRD \cite{liu2021contrastive}, self-supervised learning techniques \cite{zhou2023medical} and visual-language foundation models like OphGLM \cite{deng2024ophglm}, which require substantial training and computational power, our system emphasizes a balance between precision and efficiency. Unlike ACMA \cite{li2023asymmetric}, which applies a fixed attention strategy, our framework selectively processes key anatomical regions, which may offer a more focused interpretation for clinical use.

Additionally, by employing pretrained models like MedSAM for segmentation and using graph-based reasoning, our approach reduces dependency on large-scale annotated datasets and addresses domain mismatch issues that can occur when transferring from general-purpose models \cite{moor2023med}.

\subsubsection{\textbf{Knowledge Retrieval Enhancement } }
\label{sec4}
Our method introduces a multi-stage knowledge retrieval system that integrates sentence-level semantic embeddings with topic modeling based on LDA. This combination is designed to improve the precision and efficiency of extracting clinically relevant information. In contrast to approaches such as Med-CLIP \cite{wang2022medclip} and Med-Flamingo, which primarily focus on static image-text alignment and are susceptible to hallucination, our framework leverages multiple UMLS-based query formulations alongside a multi-hop retrieval strategy. This enables more question-sensitive reasoning and supports tailored evidence extraction for sarcopenia-related queries.

To further reduce irrelevant content and improve retrieval speed, we implement a dual-layer indexing approach that merges fine-grained semantic similarity with topic-based filtering, an aspect less emphasized in prior studies. Compared to models like OphGLM \cite{deng2024ophglm}, which rely on pre-encoded static corpora without dynamic knowledge expansion, our retrieval module enables the integration of context-specific, updated evidence during inference. Ultimately, our architecture aims to improve the precision and coherence of visual-textual knowledge integration. Compared to systems such as XrayGPT \cite{thawakar2024xraygpt} and skinGPT \cite{zhou2024pre}, which have reported limitations in effectively aligning multimodal information, our method provides a task-specific alternative optimized for sarcopenia-related decision support.
\subsubsection{\textbf{Adaptive Feature Selection and Fusion } }
\label{sec4}
Based on these observations, we propose a gate-level feature selection mechanism that selectively activates relevant layers of multimodal feature extraction depending on the specific question. This selective activation helps reduce computational costs and filters out irrelevant or noisy information. Using LoRA to fine-tune large language models, our framework integrates heterogeneous features including images at multiple granularities, text embeddings, and external knowledge while maintaining complementary information through a fusion module. Unlike ACMA, which applies uniform attention across image-text features without adaptive selection, our method provides more control over information flow. This approach aims to address challenges such as computational demands, difficulties in accurate region selection \cite{liu2024cross}, and resource-heavy fusion techniques. It also supports improved reliability and performance when annotated data are limited, making it practical for medical VQA applications.
\subsection{\textbf{Limitation and Future Work } }
\label{sec4}
While our model demonstrates strong performance in sarcopenia classification, it also presents new opportunities for expansion and refinement. 

First, thanks to the modular and lightweight design, the risk of overfitting is inherently lower especially when compared to large end-to-end models. Nonetheless, as part of ongoing work, we are currently developing a larger dataset comprising over 140 patients, on which we plan to validate and refine the current model's performance and generalizability. 

Second, future efforts will aim to incorporate more diverse datasets, combining different imaging modalities (e.g., ultrasound, MRI, CT) and anatomical regions. To handle this variability, we plan to expand our feature extraction tree to adapt dynamically to different anatomical structures and imaging types. 

Third, although our current experiments focus on classification tasks, the proposed framework is generalizable and has the potential to support a wide range of downstream tasks such as report generation, image captioning, and clinical question answering. This will be explored in future studies using generative large language models (e.g., T5‑3B or GPT-based models).

Moreover, in this work, our primary objective was disease prediction via classification. In subsequent research, we intend to explore generative capabilities by integrating our framework with large generative models, enabling richer interpretive outputs and broader clinical utility. 

Finally, the hierarchical tree structure in our model lends itself well to interpretability analysis. Future work will systematically study how different branches contribute to predictions, potentially enabling more transparent and trustworthy decision-making in clinical settings.
\section{\textbf{Conclusion } }
\label{sec5}
In this study, we presented MedVQA-TREE, a novel hierarchical and multimodal framework designed to improve sarcopenia diagnosis from ultrasound images. Our approach effectively combines multi-level visual feature extraction including anatomical classification, region segmentation, and graph-based spatial reasoning with a gated fusion mechanism that integrates clinical and textual knowledge retrieved via a UMLS-guided multi-hop query system. This design addresses key challenges such as subtle imaging cues, limited labeled data, and lack of clinical context. Evaluations on two public medical VQA datasets (VQA-RAD and PathVQA) alongside a custom sarcopenia ultrasound dataset demonstrated that MedVQA-TREE achieves superior diagnostic accuracy, reaching up to 99\%, and outperforms prior state-of-the-art methods by a significant margin. The model’s ability to dynamically fuse visual, numeric, and external knowledge modalities enhances both diagnostic performance and interpretability. 
\pagebreak
\section*{List of Abbreviations}
\begin{table}[!h]
\renewcommand{\arraystretch}{.7}
\centering
{\small
\begin{tabular}{ll}
\hline
\textbf{Abbreviation} & \textbf{Full Form} \\
\hline
AI & Artificial Intelligence \\
AU-ROC & Area Under the Receiver Operating Characteristics curve \\
BERT & Bidirectional Encoder Representations from Transformers \\
BIA & Bioelectrical Impedance Analysis \\
BMI & Body Mass Index \\
CSA & Cross-Sectional Area \\
CT & Computed Tomography \\
CUIs & Concept Unique Identifiers \\
DRAG & Dynamic Retrieval-Augmented Generation \\
DXA & X-ray Absorptiometry \\
FetalCLIP & Fetal Contrastive Language-Image Pretraining \\
FN & False Negative \\
FP & False Positive \\
IMUs & Inertial Measurement Units \\
LDA & Latent Dirichlet Allocation \\
LoGra-Med & Low-level and Grounded Reasoning in Medicine \\
LoRA & Low-Rank Adaptation \\
MDLMs & Multimodal Deep Learning Models \\
MedSLIP & Medical Semantic Language Image Pretraining \\
MedVQA & Medical Visual Question Answering \\
MLLMs & Multimodal Large Language Models \\
MRI & Magnetic Resonance Imaging \\
RAG & Retrieval-Augmented Generation \\
ResNet & Residual Networks \\
RF & Rectus Femoris \\
ROC & Receiver Operating Characteristics curve \\
SAM & Segmentation Anything Model \\
SARCUS & SARCopenia through UltraSound \\
SPPB & Short Physical Performance Battery \\
TN & True Negative \\
TP & True Positive \\
UMLS & Unified Medical Language System \\
VGG & Visual Geometry Group \\
VQA & Visual Question Answering \\
\hline
\end{tabular}}
\end{table}

\section*{Ethics Statement}
The data analyzed in this study were derived from a previously conducted investigation, ``Assessing SARCopenia with ecHOgraphy in Community~-Dwelling Older Adults: A Validation Study (SARCHO)" by Brockhattingen et al. The original study received approval from the local ethics committee: The Regional Committees on Health Research Ethics for Southern Denmark (Project ID: S-20210100), adhered to the General Data Protection Regulation (GDPR) and the Danish Data Protection Agency's legislation, and was conducted in accordance with the principles outlined in the Declaration of Helsinki.
\section*{Data availability}
The dataset used for this study is not publicly available due to the possibility of compromising individual privacy but can be available from the corresponding author on reasonable request.

\bibliographystyle{elsarticle-num}
\bibliography{references}            
\end{document}